\documentclass[letterpaper,twocolumn,10pt]{article}
\usepackage{usenix,epsfig,endnotes}

\usepackage{times}
\usepackage[font=bf]{caption}

\usepackage{mathtools}
\DeclarePairedDelimiter\ceil{\lceil}{\rceil}

\usepackage[compact]{titlesec}

\usepackage{multirow}
\usepackage[sort,nocompress]{cite}
\usepackage[usenames,dvipsnames]{color}
\usepackage{listings}
\usepackage{clrscode}

\usepackage{amsmath}
\usepackage{amsfonts}
\usepackage{times}
\usepackage{xspace} 
\usepackage{epsfig} 
\usepackage{amsmath}
\usepackage[hyphens]{url}
\usepackage{amsfonts} 
\usepackage{listings}
\usepackage{fancyvrb}
\usepackage{graphicx}
\usepackage{subfig}
\usepackage[english]{babel}
\usepackage{courier}

\usepackage{amsfonts}

\VerbatimFootnotes
 
\newcommand{\tightcaption}[1]{\vspace{-0.2cm}\caption{#1}\vspace{-0.2cm}}

\newcommand{\comment}[1]{}
\newcounter{note}[section]

\newcommand{\Section}{\S}

\usepackage{pifont}

\let\OLDthebibliography\thebibliography
\renewcommand\thebibliography[1]{
  \OLDthebibliography{#1}
  \setlength{\parskip}{0pt}
  \setlength{\itemsep}{0pt}
}

\newcommand{\Name}{Bohatei\xspace}

\newcommand{\ddos}{DDoS\xspace}

\newcommand{\learning}{learning\xspace}
\newcommand{\Learning}{Learning\xspace}

\newcommand{\mux}{load balancer\xspace}
\newcommand{\muxing}{distribution\xspace}

\newcommand{\adversary}{adversary\xspace}

\newcommand{\Cost}{\ensuremath{\mathit{Cost}}}
\newcommand{\coeffVMs}{\ensuremath{\mathit{c_1}}}
\newcommand{\coeffAttack}{\ensuremath{\mathit{c_2}}}
\newcommand{\VMs}{\ensuremath{\mathit{VMs}}}
\newcommand{\SuspiciousLegit}{\ensuremath{\mathit{Suspicious.Legit}}}
\newcommand{\UnstoppedAttack}{\ensuremath{\mathit{UnstoppedAttack}}}

\newcommand{\overprovisionFactor}{\ensuremath{\mathit{\gamma}}}

\newcommand{\epoch}{t}

\newcommand{\attackBudget}{\ensuremath{\mathit{B}}}

\newcommand{\edgepopindex}{\ensuremath{\mathit{e}}}
\newcommand{\edgepop}{\ensuremath{\mathit{E}}}
\newcommand{\attackindex}{\ensuremath{\mathit{a}}}
\newcommand{\maxAttack}{\ensuremath{\mathit{A}}}
\newcommand{\Defense}{\ensuremath{\mathit{DAG}}}
\newcommand{\AttackSet}{\ensuremath{\mathit{AttackSet}}}
\newcommand{\datacenterindex}{\ensuremath{\mathit{d}}}
\newcommand{\datacenter}{\ensuremath{\mathit{D}}}
\newcommand{\traffic}{\ensuremath{\mathit{T}}}
\newcommand{\atraffic}{\ensuremath{\mathit{t}}}
\newcommand{\vms}{\ensuremath{\mathit{n}}}
\newcommand{\capacity}{\ensuremath{\mathit{C}}}
\newcommand{\cost}{\ensuremath{\mathit{L}}}
\newcommand{\fraction}{\ensuremath{\mathit{f}}}

\newcommand{\graph}{\ensuremath{\Defense}}
\newcommand{\vertexes}{\ensuremath{\mathit{V}}}
\newcommand{\vertex}{\ensuremath{\mathit{v}}}
\newcommand{\vertexindexi}{\ensuremath{\mathit{i}}}
\newcommand{\vertexindexj}{\ensuremath{\mathit{i'}}}
\newcommand{\edges}{\ensuremath{\mathit{E}}}
\newcommand{\edge}{\ensuremath{\mathit{e}}}
\newcommand{\weight}{\ensuremath{\mathit{W}}}
\newcommand{\logical}{\ensuremath{\mathit{annotated}}}
\newcommand{\physical}{\ensuremath{\mathit{physical}}}
\newcommand{\power}{\ensuremath{\mathit{P}}}

\newcommand{\vmassigned}{\ensuremath{\mathit{q}}}
\newcommand{\vmindex}{\ensuremath{\mathit{vm}}}
\newcommand{\trafficunit}{\ensuremath{\mathit{l}}}
\newcommand{\servers}{\ensuremath{\mathit{S}}}
\newcommand{\aserver}{\ensuremath{\mathit{s}}}

\newcommand{\unitIntraRackCost}{\ensuremath{\mathit{IntraUnitCost}}}
\newcommand{\unitInterRackCost}{\ensuremath{\mathit{InterUnitCost}}}
\newcommand{\datacentercost}{\ensuremath{\mathit{dsc}}}
\newcommand{\interrack}{\ensuremath{\mathit{interR}}}
\newcommand{\intrarack}{\ensuremath{\mathit{intraR}}}
\newcommand{\maxVol}{\ensuremath{\mathit{MaxVol}}}
\newcommand{\maxVM}{\ensuremath{\mathit{MaxVM}}}

\newcommand{\classifier}{classification\xspace}

\newcommand{\node}{node\xspace}
\newcommand{\nodes}{nodes\xspace}

\newcommand{\module}{module\xspace}
\newcommand{\modules}{modules\xspace}

\newcommand{\Traffic}{\ensuremath{\mathit{T}}}

\newcommand{\mypara}[1]{\smallskip\noindent{\bf {#1}:}~}
\newcommand{\myparatight}[1]{\smallskip\noindent{\bf {#1}:}~}

\newcommand{\vyas}[1]{{\footnotesize\color{red}[VS: #1]}}
\newcommand{\seyed}[1]{{\footnotesize\color{blue}[SF: #1]}}
\newcommand{\bailey}[1]{}
\newcommand{\shepherd}[1]{}

\newcounter{packednmbr}

\newenvironment{packedenumerate}{\begin{list}{\thepackednmbr.}{\usecounter{packednmbr}\setlength{\itemsep}{0.5pt}\addtolength{\labelwidth}{0pt}\setlength{\leftmargin}{\labelwidth}\setlength{\listparindent}{\parindent}\setlength{\parsep}{3pt}\setlength{\topsep}{0pt}}}{\end{list}}

\newenvironment{packeditemize}{\begin{list}{$\bullet$}{\setlength{\itemsep}{0.5pt}\addtolength{\labelwidth}{0pt}\setlength{\leftmargin}{\labelwidth}\setlength{\listparindent}{\parindent}\setlength{\parsep}{3pt}\setlength{\topsep}{0pt}}}{\end{list}}

\begin{document}

\title{\Name: Flexible and Elastic \ddos Defense}

\author{Seyed K. Fayaz, Yoshiaki Tobioka, Vyas Sekar, Michael Bailey}

\author{Seyed K. Fayaz$^*$  \hspace{4mm} Yoshiaki Tobioka$^*$  \hspace{4mm} Vyas Sekar$^*$  \hspace{4mm} Michael Bailey$^\dagger$ 
\\ $^*$CMU \hspace{5 mm} $^\dag$UIUC} 

\maketitle 

\begin{abstract}
\ddos defense today relies on expensive and proprietary hardware appliances 
deployed at fixed locations.  This introduces key limitations 
with respect to flexibility (e.g., complex routing to get traffic to these 
``chokepoints'') and elasticity in handling changing attack patterns.  We 
observe an opportunity to address these limitations using new networking paradigms 
such as software-defined networking (SDN) and network functions virtualization 
(NFV).  Based on this observation, we design and implement  
\Name, a  {\em flexible} and {\em   elastic} \ddos defense system. In designing 
\Name, we   address key challenges with respect to scalability, responsiveness, 
and   adversary-resilience. We have implemented defenses for several \ddos 
attacks using \Name.  Our evaluations show that \Name is scalable (handling 
500 Gbps attacks), responsive (mitigating attacks within one minute), and 
resilient to dynamic adversaries.

\end{abstract}

\section{Introduction}
\label{sec:intro}

In spite of extensive industrial and academic efforts
(e.g.,~\cite{reihersurvey, Moore:2006:IID:1132026.1132027,
arbor}), distributed denial-of-service
(\ddos) attacks continue to plague the Internet. Over the last few
years, we have observed a dramatic escalation in the number, scale,
and diversity of \ddos attacks.  For instance, recent estimates
 suggest that over 20,000 \ddos attacks occur per day~\cite{arbor-number},
 with peak volumes of 0.5~Tbps~\cite{hockey_stick_era, gorilla}. At the same time, new
vectors~\cite{core melt, crossfire} and variations of known
attacks~\cite{ntp_hell} are constantly emerging. The
damage that these \ddos attacks cause to organizations is well-known and include both
  monetary losses (e.g., \$40,000 per hour~\cite{incapsula})
 and loss of customer trust.


 \ddos defense today is implemented using expensive and proprietary 
hardware appliances (deployed in-house or in the 
cloud~\cite{prolexic, cloudflare}) that are \emph{fixed} in terms 
of placement, functionality, and capacity. First, they are typically
deployed at fixed network aggregation points (e.g., a peering edge 
link of an ISP). Second, they provide fixed functionality with respect 
to the types of \ddos attacks they can handle. Third, they have a 
fixed capacity with respect to the maximum volume of traffic 
they can process. This fixed nature of today's approach 
leaves network operators with two unpleasant options: (1) to 
overprovision by deploying defense appliances 
that can handle a high (but pre-defined) volume of every known 
attack type at each of the aggregation points, or (2) to deploy
a smaller number of defense appliances at a central location (e.g., 
a scrubbing center) and reroute traffic  to  this location.  While 
option (2) might be more cost-effective, it raises two other challenges. 
First, operators run the risk of underprovisioning. Second, 
traffic needs to be explicitly routed through a fixed central location, 
which introduces additional traffic latency and requires complex 
routing hacks (e.g.,~\cite{irscp-usenix07}). Either way, handling larger 
volumes or new types of attacks typically mandates purchasing and 
deploying new hardware appliances.

\comment{
 Today, detection and mitigation of  \ddos attacks is implemented via proprietary
hardware appliances.  These appliances may be hosted in-house  or outsourced to
cloud-based services~\cite{prolexic, cloudflare}.  In either case, there
are two natural placement and provisioning options.  The first option is to place
 appliances  at natural network aggregation points  and provision
them for the link capacity at each location. However, this leads to significant
overprovisioning---each device is typically scaled to meet the worst case
attack scenario at each location (e.g., the entire link bandwidth) and devices
are placed everywhere along the aggregation frontier (e.g., all of the peering
edge links) in anticipation of arbitrary attack sources and destinations. The
second option is to consolidate some number of devices at a fixed location
(e.g., a scrubbing center) and reroute traffic  to  these
devices.  While this is  more cost-effective, it raises two other challenges.
First, the operators run the risk of underprovisioning and not being able to
fully mitigate a given attack. Second, traffic needs to be explicitly routed to
a fixed location. This introduces additional traffic latency and also requires
complex routing hacks (e.g.,~\cite{irscp-usenix07}).
}

 Ideally, a DDoS defense architecture should provide the {\em
flexibility} to seamlessly place defense mechanisms where they are 
needed and the {\em elasticity} to launch defenses as needed depending on the type and scale
of the attack. We observe that similar problems in other areas of network management 
have been tackled by taking advantage of two new  paradigms: software-defined 
networking (SDN)~\cite{4d, openflow} and network functions
virtualization (NFV)~\cite{etsinfv}. SDN simplifies routing by
decoupling the control plane (i.e., routing policy) from the data plane (i.e., switches).  In
parallel, the use of virtualized network functions via NFV reduces cost and
enables elastic scaling and reduced time-to-deploy  akin to cloud
computing~\cite{etsinfv}. These potential benefits have led major industry
players (e.g., Verizon, AT\&T) to embrace SDN and NFV~\cite{att_vision, att_intel,
verizon_sdn,ons_14_keynote}.\footnote{To quote the SEVP of AT\&T: ``To say 
that we are both feet in [on SDN] would be an
understatement. We are literally all in~\cite{att_intel}.''}

In this paper, we present \Name\footnote{It means breakwater in Japanese, used 
to defend against tsunamis.}, a flexible and \mbox{elastic}  \ddos defense system that
demonstrates the benefits of these new network management paradigms in the
context of \ddos defense. \Name leverages NFV capabilities to elastically vary the
required scale (e.g., 10 Gbps vs.\ 100 Gbps attacks) and type (e.g.,
SYN proxy vs.\ DNS reflector defense) of \ddos defense realized by defense
virtual machines (VMs). Using the flexibility of SDN, \Name steers suspicious 
traffic through the defense VMs while minimizing user-perceived latency and network 
congestion.

In designing \Name, we address three  key algorithmic and system design
challenges. First, the resource management problem to determine 
the number and location of defense VMs is NP-hard and takes hours
to solve. Second, existing SDN solutions are fundamentally unsuitable for
\ddos defense (and even introduce new attack avenues) because they rely 
on a per-flow orchestration paradigm, where switches need to contact a network 
controller each time they receive a new flow. Finally, an intelligent \ddos adversary 
can attempt to evade an elastic defense, or alternatively induce provisioning 
inefficiencies by dynamically changing attack patterns.

We have implemented a \Name controller using  {\tt OpenDaylight}~\cite{odl}, an
industry-grade SDN platform. We have  used a combination of open source tools (e.g.,
OpenvSwitch~\cite{openvswitch}, Snort~\cite{snort}, Bro~\cite{bro},
iptables~\cite{iptables}) as defense modules. We have developed a scalable 
resource management algorithm. Our
evaluation, performed on a real testbed as well as using simulations, shows
that \Name effectively defends against several different \ddos attack types,
scales to scenarios involving 500 Gbps attacks and ISPs with about 200
backbone routers, and can effectively cope with dynamic adversaries.

\mypara{Contributions and roadmap} In summary, this paper makes the following 
 contributions: 
\begin{packeditemize}
\item Identifying new opportunities via SDN/NFV to improve
the current \ddos defense practice (\Section\ref{sec:motivation});

\item Highlighting the challenges of  applying  existing  SDN/NFV 
techniques in the context of \ddos defense(\Section\ref{sec:vision}); 

\item Designing a  responsive resource management algorithm   
  that is 4-5  orders of magnitude faster than the state-of-the-art solvers 
  (\Section\ref{sec:resource_layer}); 
  
\item Engineering a practical and scalable network orchestration mechanism 
using proactive tag-based forwarding 
 that avoids the pitfalls of existing SDN solutions (\Section\ref{sec:orchestration_layer});
 
\item An adaptation strategy to handle dynamic adversaries
that can change the \ddos attack mix over time (\Section\ref{sec:policy_layer});
 
\item A proof-of-concept implementation to handle several known \ddos 
 attack types using industry-grade SDN/NFV platforms (\Section\ref{sec:implementation}); and 

\item  A systematic demonstration of the scalability and effectiveness of \Name
  (\Section\ref{sec:evaluation}). 
  
\end{packeditemize}

We discuss related work (\Section\ref{sec:relwork}) before concluding (\Section\ref{sec:conc}).


\section{Background and Motivation}
\label{sec:motivation}

In this section, we give a brief overview of software-defined networking (SDN)
and network functions virtualization (NFV) and discuss new  opportunities these
 can enable in the context of \ddos defense. 

\comment{
 \begin{figure*}[t]
\begin{center}
\subfloat[Elasticity with respect to attack volume.]
{
 \includegraphics[width=133pt]{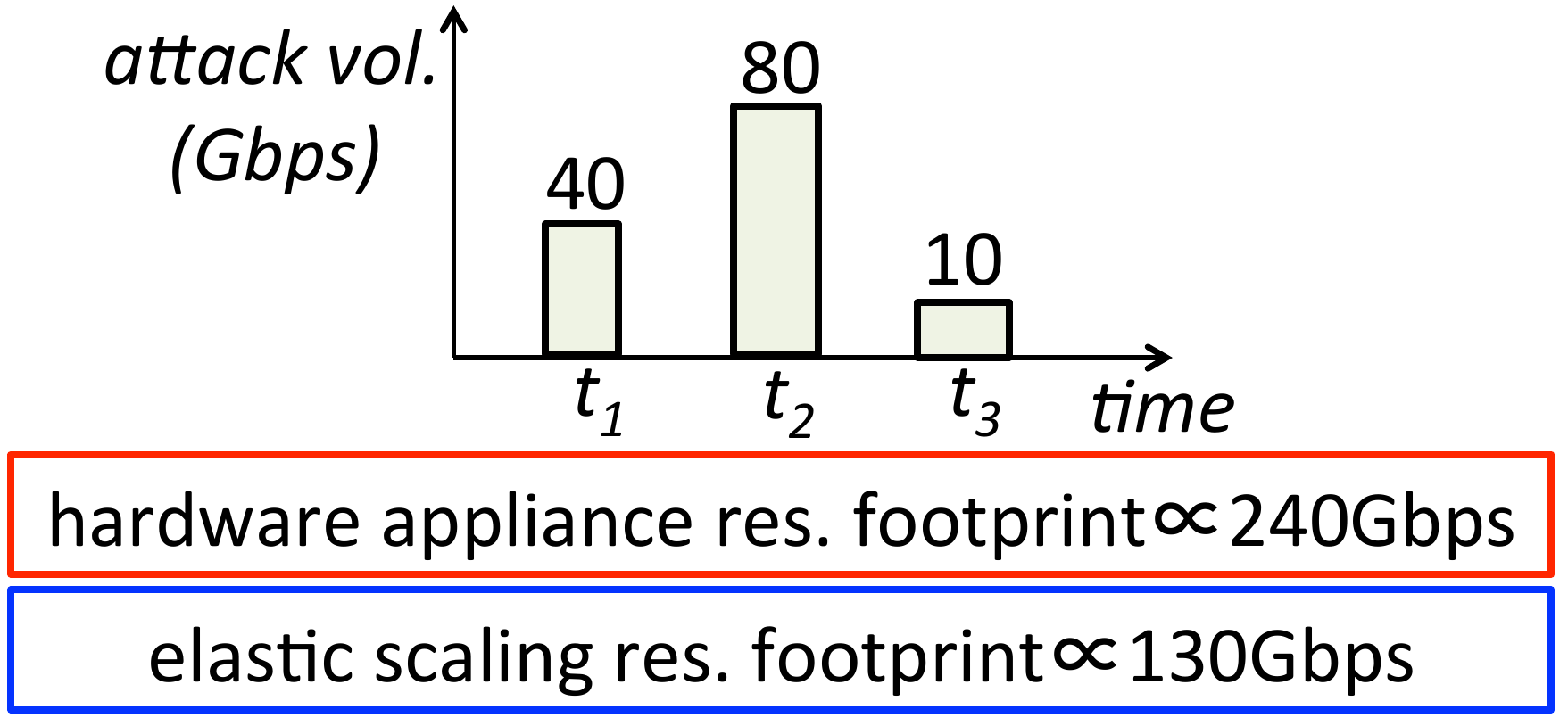}\hspace{0em}
\label{fig:bad_resource}
}
\subfloat[Flexibility with respect to attack types.]
{
 \includegraphics[width=145pt]{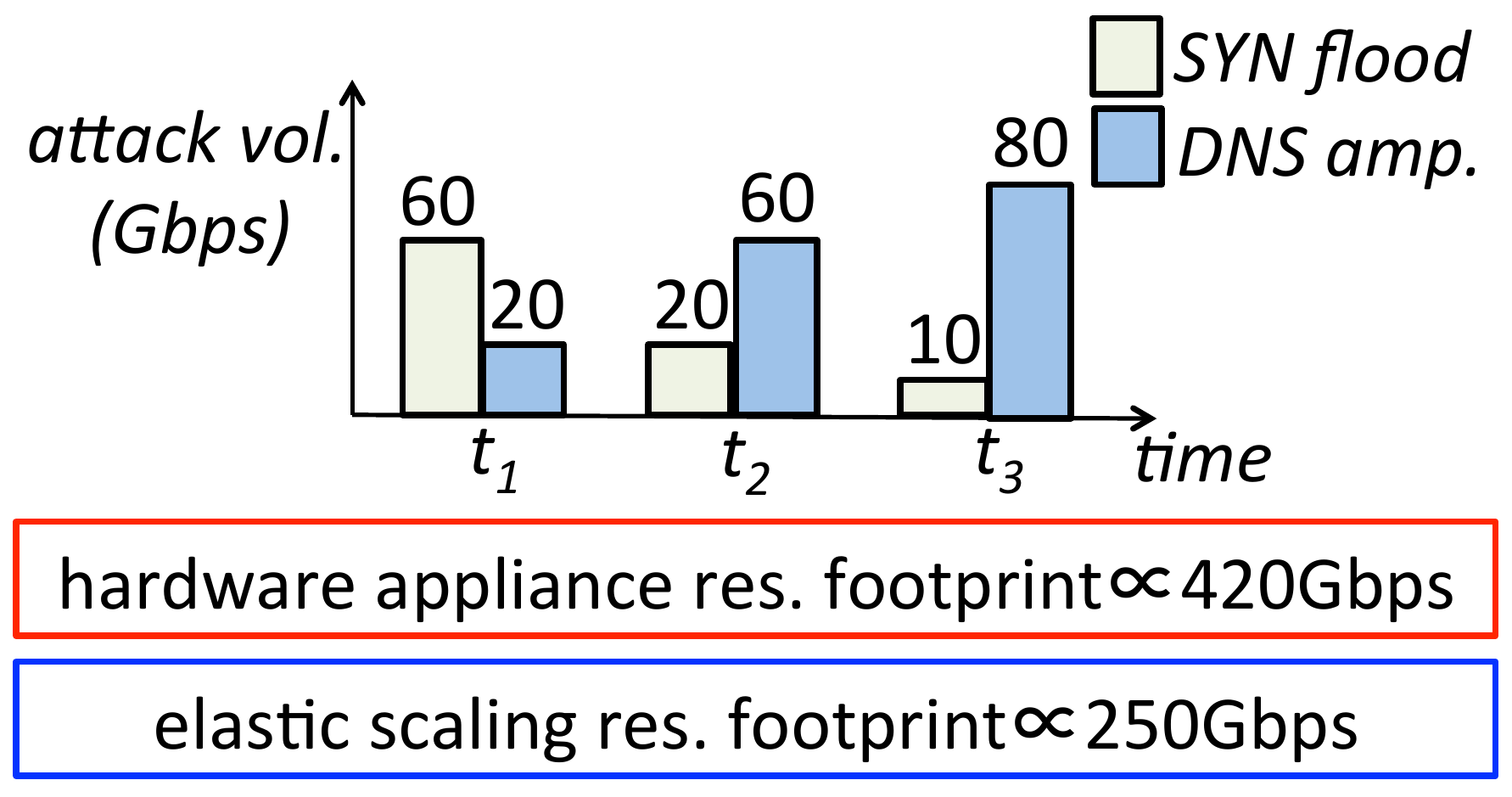}
\label{fig:bad_attack_type}
}
\subfloat[Routing efficiency.]
{
 \includegraphics[width=185pt]{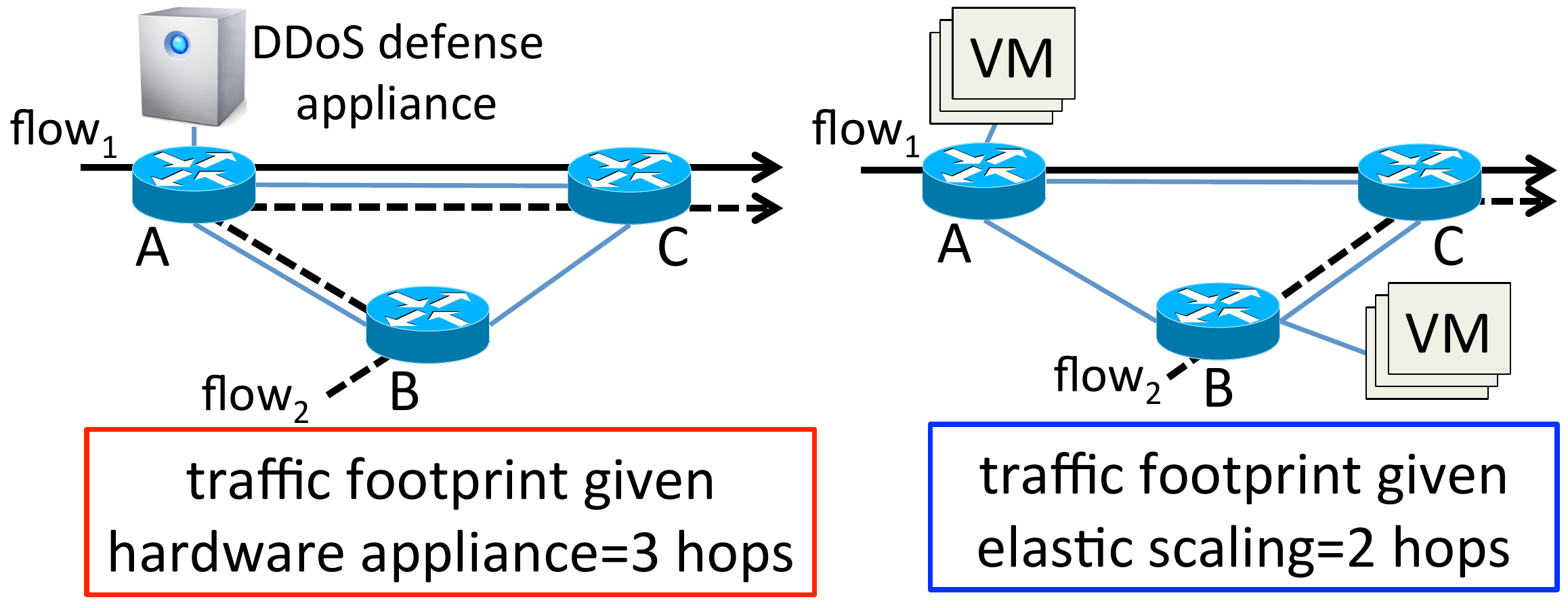}\hspace{0em}
\label{fig:bad_routing}
}
\end{center}
 \vspace{-0.3cm}
\tightcaption{New opportunities in \ddos defense. ($\propto$ represents ``proportional to''.)}
\label{fig:today}
 \vspace{-0.3cm}
 \end{figure*}
}

\subsection{New network management trends}
\label{subsec:background}
\mypara{Software-defined networking (SDN)} Traditionally, network 
 control  tasks (e.g., routing, traffic engineering, and access 
control)  have been tightly coupled with  their data plane 
 implementations (e.g., distributed routing protocols, ad hoc ACLs). 
This practice has made network management complex,
brittle, and error-prone~\cite{4d}. SDN simplifies 
network management  by decoupling the  
network control plane (e.g., an intended routing policy) from the network 
data plane (e.g., packet forwarding by individual switches). Using SDN, 
a network operator can centrally program the network behavior through 
APIs such as OpenFlow~\cite{openflow}.  This flexibility has motivated 
several real world deployments to transition to SDN-based 
 architectures (e.g., \cite{google_sdn}).

\mypara{Network functions virtualization (NFV)} Today, network functions (e.g.,
firewalls, IDSes) are  implemented using specialized hardware. While this 
practice was necessary for performance reasons, it leads to high 
cost and inflexibility. These limitations have motivated the use of virtual 
network functions  (e.g., a virtual firewall) on general-purpose
servers~\cite{etsinfv}. Similar to traditional virtualization, NFV reduces 
costs and enables new opportunities (e.g., elastic scaling).  Indeed, leading  
vendors already offer virtual appliance products (e.g.,~\cite{zscalar}).
\shepherd{C-2}
Given these benefits, major ISPs have deployed (or are planning to  deploy)
datacenters to run virtualized 
functions that replace  existing specialized
hardware~\cite{att_vision, verizon_sdn,ons_14_keynote}.  One
potential concern with NFV  is low packet processing performance. Fortunately,
several recent advances enable line-rate (e.g., 10-40Gbps) packet
processing by software running on commodity hardware~\cite{arrakis}.  
Thus, such performance concerns are  increasingly a non-issue and will 
further diminish given constantly improving hardware support~\cite{dpdk}.
 

\subsection{New opportunities in \ddos defense}
\label{subsec:opportunities}

\shepherd{A-4, A-5:This subsection has been substantially improved to make
the benefits of SDN/NFV in \ddos defense clear.}

Next, we briefly highlight new opportunities that  SDN and NFV can enable 
 for \ddos defense. 

\comment
{
 \begin{table}[t]
  \begin{center}
  \begin{footnotesize}
  \begin{tabular}{p{2.2cm}|p{2.4cm}}
         Throughput (Gbps)     & Starting price	\\ \hline
         1              	 &  \$11,000 - \$38,000  	\\ \hline
         2              	 &  \$20,000 - \$63,000  	\\ \hline
	3              	 &  \$54,000  	\\ \hline
	4              	 &  \$68,000  	\\ \hline
	8              	 &  \$98,000 - \$112,000  	\\ \hline 
	12              &  \$128,000  	\\  
     \end{tabular}
  \end{footnotesize}
  \end{center}
  \vspace{-0.3cm}
 \tightcaption{Price range for \ddos appliances~\cite{gsa_schedules}.}
  \vspace{-0.3cm}
 \label{table:cost}
 \end{table}
}

\mypara{Lower capital costs} Current \ddos defense is based on specialized
hardware appliances (e.g.,~\cite{arbor,radware}).  Network  operators either
deploy them on-premises, or outsource \ddos defense to a remote packet
scrubbing site (e.g.,~\cite{cloudflare}).  In either  case, \ddos defense is 
expensive. For instance, based on public
estimates from the  General Services Administration (GSA) Schedule, a 10~Gbps
\ddos defense appliance costs $\approx$\$128,000~\cite{gsa_schedules}.  To put
this in context, a commodity server with a 10~Gbps Network Interface Card (NIC)
costs about \$3,000~\cite{dell_server}. This suggests roughly 1-2 orders of
magnitude potential reduction in capital expenses  (ignoring software and
development costs) by moving from specialized appliances  to commodity
hardware.\footnote{\shepherd{A-9} Operational expenses are harder to 
compare due to the lack of publicly available data.}

\mypara{Time to market} As new and larger attacks emerge, enterprises today
need to frequently purchase more capable hardware appliances and 
integrate them into the network infrastructure. This is an expensive and tedious 
process~\cite{etsinfv}. In contrast, launching a VM customized for a new type 
of attack, or  launching more VMs to handle larger-scale attacks, is trivial using 
SDN and NFV.



\mypara{Elasticity with respect to attack volume} Today, 
  \ddos defense  appliances deployed at network chokepoints need 
to be provisioned  to handle a predefined maximum attack volume. 
As an illustrative example, consider an enterprise network where a \ddos 
scrubber appliance is deployed at each ingress point.  Suppose 
the projected resource footprint (i.e., defense resource usage over time)
to defend against a SYN flood attack at times $t_1$, $t_2$, and 
$t_3$ is 40, 80, and 10 Gbps, respectively.\footnote{For brevity, we use the traffic 
volume as a proxy for the memory consumption  and CPU cycles required 
to handle the traffic.} The total 
resource footprint over this entire time period  is 
$3 \times max \{40,80,10\}=240$ Gbps, as we need to provision 
for the worst case.
 However, if we could elastically scale the defense capacity, we
would only introduce a resource footprint of $40+80+10=130$ Gbps---a 
45\% reduction in defense resource footprint. This reduced hardware 
footprint can  yield energy savings and allow ISPs to repurpose the 
hardware for other services.  


\mypara{Flexibility with respect to attack types} 
Building on the above example, suppose in
addition to the SYN flood attack, the projected resource footprint for a DNS 
amplification attack in time intervals $t_1$, $t_2$, and $t_3$ is 
20, 40, and 80 Gbps, respectively. 
Launching only the required types of defense VMs as opposed to using monolithic
appliances (which handle both attacks), drops the hardware footprint by $40\%$;
i.e., from $3 \times (max\{40,80,10\} + max\{20,40,80\})=480$ to $270$.

\mypara{Flexibility with respect to vendors} 
Today, network operators are locked-in to the defense capabilities offered by 
specific vendors.  In contrast,  with SDN and  NFV, they can launch appropriate 
best-of-breed defenses. For example, suppose vendor 1 is better for 
SYN flood defense, but vendor 2 is better for DNS flood defense. The 
physical constraints today may force an ISP to pick only one hardware appliance. With 
SDN/NFV we can avoid the undesirable situation of picking
only one vendor and rather have a deployment with both types of VMs each 
for a certain type of attack. Looking even further, we  also envision that network
operators can mix and match capabilities  from different vendors; e.g., if
vendor 1 has better detection capabilities but vendor 2's blocking algorithm is
more effective, then we can flexibly combine these two to create a more 
powerful defense platform. 
 
 \begin{figure}[t]
\begin{center}
\includegraphics[width=190pt]{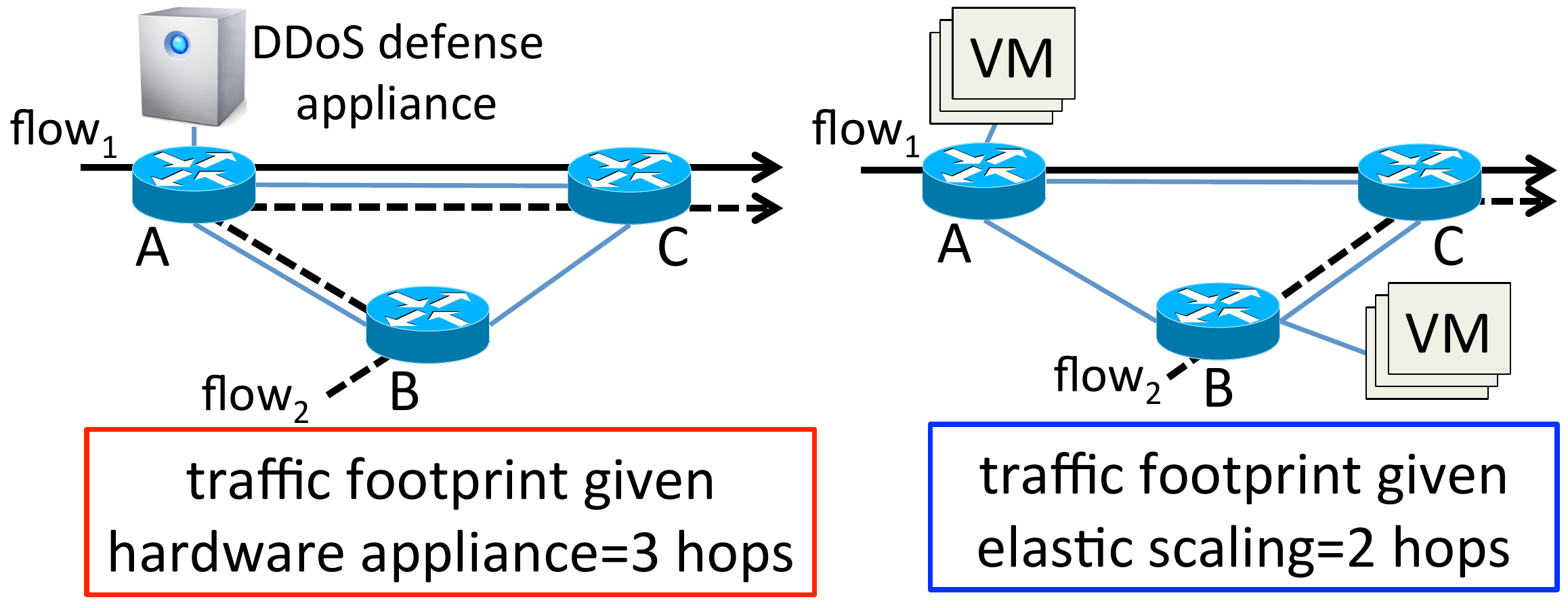}
\vspace{-0.1cm}
\tightcaption{\ddos defense routing efficiency enabled by  SDN and NFV.}
\label{fig:bad_routing}
\vspace{-0.6cm}
\end{center}
\end{figure}

\mypara{Simplified and efficient routing} \shepherd{A-8} Network operators 
today need to  employ complex routing hacks (e.g.,~\cite{irscp-usenix07}) to steer traffic 
through a fixed-location \ddos hardware appliance (deployed either 
on-premises or in a remote site). As Figure~\ref{fig:bad_routing} illustrates, this
causes additional latency.  Consider two end-to-end flows 
$flow_1$ and $flow_2$. Way-pointing $flow_2$ through the appliance 
(the left hand side of the figure) makes the total path lengths 3 hops. But if we 
could launch VMs where they are needed (the right hand side of the figure), 
we could drop the total path lengths to 2 hops---a 33\% decrease in traffic footprint. 
Using NFV we can launch defense VMs on the closest location to where they 
are currently needed, and using SDN we can flexibly route traffic 
through them.

In summary,  we observe new opportunities to build a flexible and elastic \ddos
defense mechanism via SDN/NFV. In the next section, we highlight the challenges 
in realizing these benefits.



\section{System Overview}
\label{sec:vision}


In  this section, we envision the deployment model and
workflow of  \Name, highlight the challenges in realizing our 
vision, and outline our key ideas to address these challenges.

\subsection{Problem scope}
\label{subsec:scope}
 
\mypara{Deployment scenario} For concreteness, we focus on an ISP-centric
deployment model, where an   ISP offers \ddos-defense-as-a-service to its
customers. Note that several ISPs already have such
commercial offerings (e.g.,~\cite{attprotect}).
  We envision different monetization 
avenues. For example, an ISP can offer a value-added security service to its 
customers that can replace the customers' in-house \ddos defense hardware. Alternatively, the 
ISP can allow its customers to use \Name as a cloudbursting option when 
the attack exceeds the customers' on-premise hardware.  While we describe 
our work in an ISP setting,  our ideas are general and can be applied to 
other deployment models; e.g., CDN-based \ddos defense or deployments 
inside cloud providers~\cite{prolexic}. 


\begin{figure}[t]
\begin{center}
\includegraphics[width=220pt]{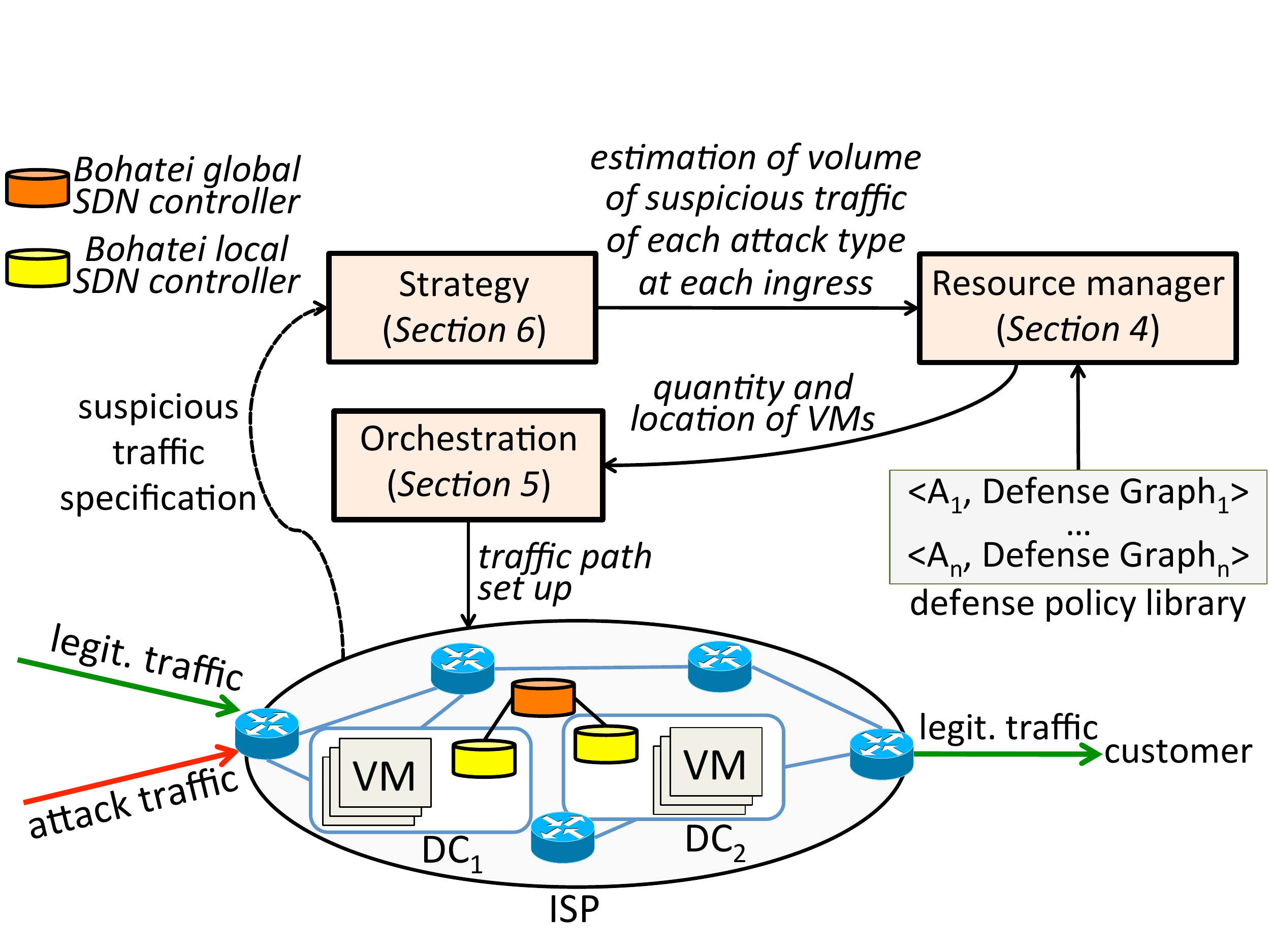}
\vspace{-0.2cm}
\tightcaption{\Name system overview and workflow.}
\label{fig:overview}
\vspace{-0.4cm}
\end{center}
\end{figure}

In addition to traditional backbone routers and interconnecting links, we
envision the ISP has deployed multiple datacenters as shown in
Figure~\ref{fig:overview}. Note that this is not a new requirement; ISPs 
already have several in-network datacenters and are planning  additional 
rollouts in the near future~\cite{verizon_sdn, ons_14_keynote}.  Each 
datacenter has commodity hardware servers and can run standard 
virtualized network functions~\cite{ananta}.

\bailey{many customers are hybrid, both in house deployments with
  cloud ``insurance policies''}

\shepherd{A-1: The next two paragraphs makes the use case of \Name clear}

\mypara{Threat model} We focus on a general \ddos threat against the 
victim, who is a  customer of the ISP. The adversary's aim is to exhaust 
the network bandwidth of the victim. The adversary can flexibly 
choose from a {\em set of candidate attacks} 
$\AttackSet =\{\maxAttack_{\attackindex}\}_\attackindex$. As a concrete starting point, we
consider the following types of \ddos attacks: TCP SYN flood,  UDP flood, DNS
amplification, and elephant flow.  We assume  the \adversary controls a large
number of bots, but the total \emph{budget} in terms of the maximum volume 
of attack traffic it can launch at any given time is fixed. Given the budget, the
adversary has a complete control over the choice of (1) type and  mix of attacks
from the $\AttackSet$ (e.g., 60\% SYN and 40\% DNS) and (2) the set of ISP ingress
locations at which the attack traffic enters the ISP.  For instance, a simple
adversary may launch a single fixed attack $\maxAttack_\attackindex$  arriving at a
single ingress, while an advanced adversary may choose a mix of various attack 
types and multiple ingresses.  For clarity, we restrict our presentation to 
focus on a single customer noting that it is straightforward to extend 
our  design to support multiple customers.







\begin{figure}[t]
\begin{center}
\vspace{-0.1cm}
\includegraphics[width=160pt]{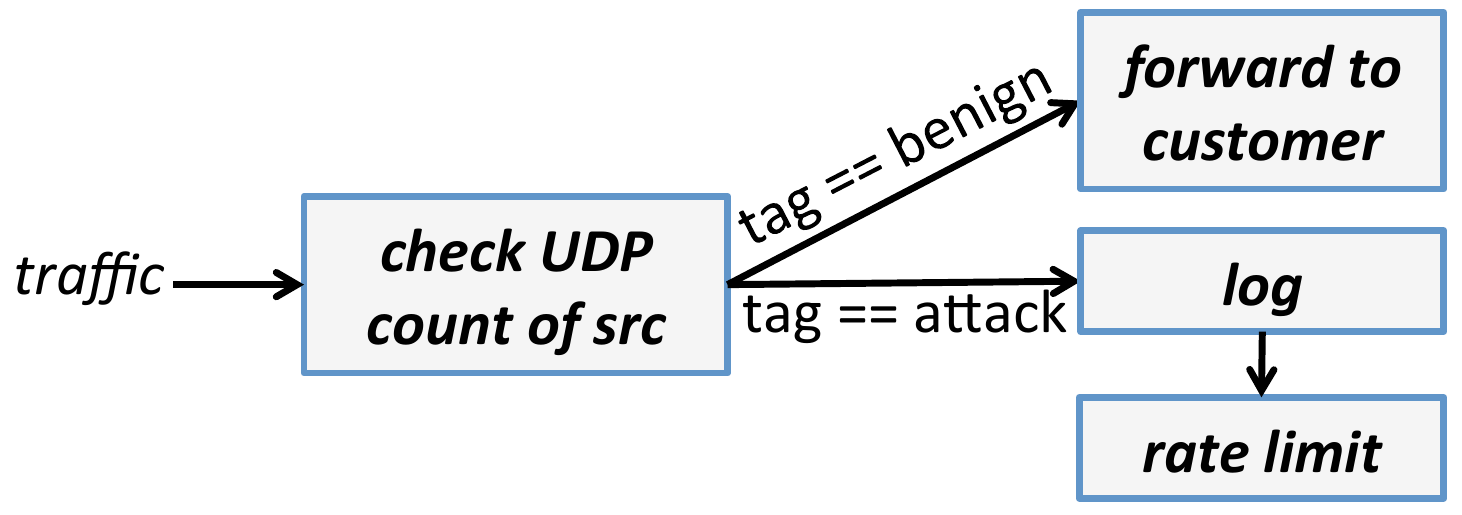}
\vspace{-0.2cm}
\tightcaption{A sample defense against UDP flood.}
\vspace{-0.8cm}
\label{example_graph}
\end{center}
\end{figure}

\mypara{Defenses} We assume the ISP has a pre-defined {\em library of defenses}
specifying a defense strategy for each attack type. For each attack type 
$\maxAttack_\attackindex$, the defense strategy is specified as 
a directed acyclic graph $\Defense_\attackindex$ representing a typical  
multi-stage attack analysis and mitigation procedure.    Each node of the graph represents a 
logical module and the edges are  tagged with the result of the previous nodes 
processing (e.g., ``benign'' or ``attack'' or ``analyze further'').   
Each  logical node will be realized by one (or more) {\em virtual appliance(s)} 
depending on the attack volume.
Figure~\ref{example_graph} shows an example strategy graph with 4 modules 
used for defending against a UDP flood attack. Here, the first module tracks the
number of UDP packets each source sends  and performs a simple threshold-based
check to decide whether the source needs to be let through or throttled.

 Our goal here is  not to develop new defense algorithms  but to develop 
 the system orchestration capabilities to enable flexible and elastic 
defense. As such, we assume the $\Defense$s have been provided by domain 
experts,  \ddos defense vendors, or by consulting 
best practices. 



\subsection{\Name workflow and challenges}
\label{subsec:challenges}

The workflow of \Name has four steps (see Figure~\ref{fig:overview}):

\begin{packedenumerate}

\item {\em Attack detection:} We assume the ISP uses some out-of-band  anomaly detection 
technique to flag whether a customer is under a \ddos attack~\cite{wavelet}. The 
design of this detection algorithm is outside the scope of this paper. The detection 
algorithm gives a coarse-grained specification of the suspicious traffic, indicating the 
customer under attack and some coarse identifications of the type and sources of the
attack; e.g., ``srcprefix=*,dstprefix=cust,type=SYN''.

\item {\em Attack estimation:} Once suspicious traffic is detected,  the strategy 
module  estimates  the volume of suspicious traffic of each attack type arriving 
at each ingress.  

\item {\em Resource management:} The  resource  manager then uses these
estimates as well as the library of defenses to determine the type, number, and the
location of defense VMs that need to be instantiated. The goal of 
the resource manager is to efficiently assign available network resources to the defense
while minimizing user-perceived latency and network congestion.  

\item {\em Network orchestration:} Finally, the  network orchestration module 
sets up the required network forwarding rules to steer suspicious traffic 
to the defense VMs as mandated by the resource manager. 

\end{packedenumerate}

Given this workflow, we highlight the three challenges we need to 
address to realize our vision:

\mypara{C1. Responsive resource management} We need an efficient way 
of assigning the ISP's available compute and network resources to \ddos
defense. Specifically, we need to decide  how many VMs of each type to run 
on each server of each datacenter location so that attack traffic is handled 
properly while minimizing the latency experienced by legitimate traffic. Doing so 
in a \emph{responsive} manner (e.g., within tens of seconds), however, is
challenging. Specifically, this entails solving a large NP-hard optimization 
problem, which can take several  hours to solve even with state-of-the-art 
solvers.

\mypara{C2. Scalable network orchestration} The canonical view in SDN  is 
to set up switch forwarding rules  in a {\em per-flow} and {\em reactive} 
manner~\cite{openflow}. That is, every  time a switch receives a flow for 
which it does not have a forwarding entry, the switch queries the  SDN controller 
to get the forwarding rule.  Unfortunately, this per-flow and reactive paradigm is 
fundamentally unsuitable for  \ddos defense. First, an adversary can easily saturate 
the control plane bandwidth as well as the controller compute 
resources~\cite{avantguard}.  Second, installing per-flow rules on the switches will  
quickly exhaust the limited rule space ($\approx$4K TCAM rules). Note that unlike 
traffic engineering applications of SDN~\cite{google_sdn}, coarse-grained IP  
prefix-based forwarding policies would not suffice in the context of \ddos 
defense, as we cannot predict the IP prefixes of future attack traffic.

\mypara{C3. Dynamic adversaries} Consider a dynamic adversary who  can rapidly
change the attack mix (i.e., attack type, volume, and ingress point).  This behavior
can make the ISP choose between two undesirable choices: (1) wasting compute 
resources by overprovisioning for attack scenarios that may not ever arrive, (2)
not instantiating the required defenses (to save resources), which will let 
attack traffic reach the customer.

\subsection{High-level approach}

Next we highlight our key ideas  to address C1--C3: 

\begin{packeditemize}

\item {\bf Hierarchical optimization decomposition
(\Section\ref{sec:resource_layer}):} To address C1, we use a hierarchical 
decomposition of the resource optimization problem into  two stages.  First, 
the \Name global  (i.e., ISP-wide) controller uses coarse-grained 
information (e.g., total spare capacity of each datacenter) to determine 
how many and what types of VMs to run in each datacenter. Then, each 
local (i.e., per-datacenter) controller  uses  more fine-grained information 
(e.g., location of available  servers) to determine the specific server 
on which each defense VM will run.

\comment{
\seyed{should we say this in this section: to minimize  the damage due 
to inefficient scaling, we exploit the modularity  of the defense graphs 
to enable fine-grained elastic scaling at  the module granularity  rather 
than a coarse-grained ``graph''  granularity.} \vyas{too low level. drop.}
}

\item {\bf Proactive tag-based forwarding
(\Section\ref{sec:orchestration_layer}):} To address C2, we design a 
scalable  orchestration mechanism using two key ideas. First, switch 
forwarding rules are based on per-VM tags rather than per-flow to 
dramatically reduce the size of the forwarding tables. Second, we 
proactively configure the switches to eliminate frequent interactions 
between the switches and the control plane~\cite{openflow}. 


\item {\bf Online adaptation (\Section\ref{sec:policy_layer}):} To handle 
a dynamic adversary that changes the attack mix (C3), we design a defense 
strategy adaptation approach inspired by classical online algorithms for 
regret minimization~\cite{kalai}.

\bailey{so this needs to be more explicit. i saw this in 2 and now again here
and its still not clear. my take is your are doing fine grain load
balancing...maybe more resources for the SYN module, and   less for the DNS
flooding module}

\end{packeditemize}


\section{Resource Manager}
\label{sec:resource_layer}

 
 The goal of the resource management module is to efficiently determine 
 network and compute resources to  analyze and take action
 on suspicious  traffic.  The key here is responsiveness---a  slow algorithm 
 enables adversaries to nullify the defense by rapidly changing their attack 
 characteristics.   In  this 
 section, we describe the  optimization  problem that \Name needs to solve and 
 then present a scalable heuristic that achieves near optimal results.


\subsection{Problem inputs} 

Before we describe the resource management problem, we establish  the main
input parameters: the ISP's compute and network parameters and the defense 
processing requirements of traffic of different attack types. We consider 
an  ISP composed of  a set of edge 
PoPs\footnote{We use the terms ``edge PoP'' and ``ingress'' interchangeably.}
$\boldsymbol{\edgepop}= \{\edgepop_{\edgepopindex}\}_\edgepopindex$ and a set of
datacenters $\boldsymbol{\datacenter}=
\{\datacenter_{\datacenterindex}\}_\datacenterindex$.

\begin{figure}
\includegraphics[width=200pt]{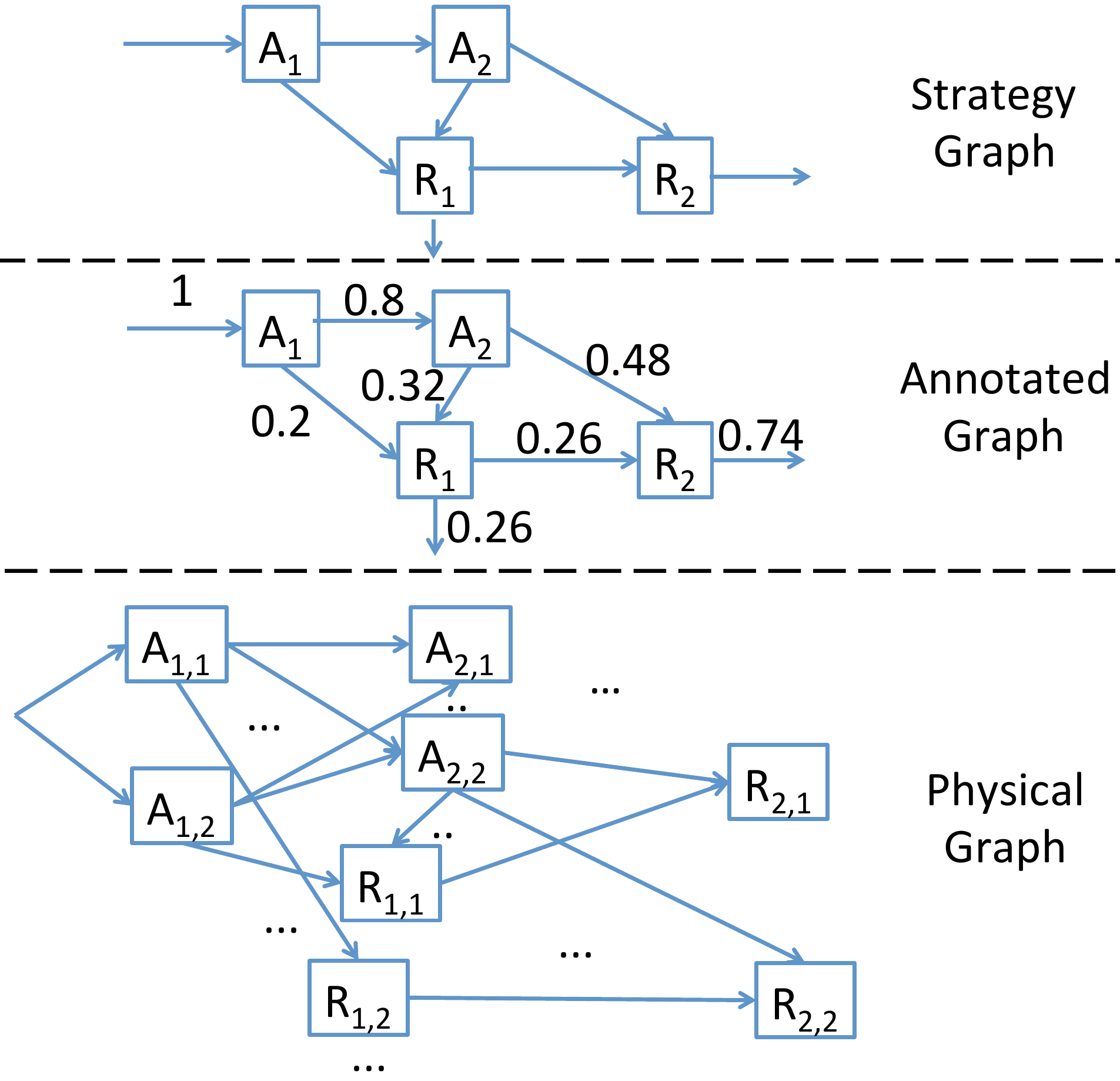}
\tightcaption{An illustration of strategy vs.\ annotated vs.\ physical graphs.
Given annotated graphs and suspicious traffic volumes, the resource 
manager computes physical graphs.}
\label{fig:resource} \end{figure}

\mypara{ISP constraints}  Each datacenter's  traffic
processing capacity  is determined by a pre-provisioned uplink capacity
$\capacity^{link}_{\datacenterindex}$  and   compute capacity
$\capacity^{compute}_{\datacenterindex}$. The compute capacity is specified in
terms of the number of VM slots, where  each VM slot has a given capacity specification (e.g., 
instance sizes in EC2~\cite{ec2}).

\mypara{Processing requirements} As discussed earlier in \Section\ref{subsec:scope}, 
different attacks require different  {\em strategy graphs}.  However, the notion 
of a strategy graph by itself will not suffice for resource management, as 
it is does not specify the traffic volume that at each \module should process.

The input to the resource manager is in form of  {\em annotated graphs} as shown 
in Figure~\ref{fig:resource}. An annotated  graph $\graph^{\logical}_\attackindex$ is 
a strategy graph annotated with edge weights, where each weight represents 
the fraction of the total input traffic to the graph that is  expected to traverse 
the corresponding edge. These weights are pre-computed based on prior network 
monitoring data (e.g., using NetFlow) and from our adaptation module 
(\Section\ref{sec:policy_layer}). 
$\Traffic_{\edgepopindex,\attackindex}$ denotes the volume of  suspicious
traffic of type $\attackindex$ arriving at  edge PoP   $\edgepopindex$. For example, 
in Figure~\ref{fig:resource}, weight 0.48 from node $A_2$ to node $R_2$ 
means 48\% of the total input traffic to the graph (i.e., to $A_1$) is expected 
to traverse edge $A_2 \rightarrow R_2$. 


Since modules may vary in terms of compute complexity and the traffic 
rate that can be handled per VM-slot, we need to account for the parameter
$\power_{\attackindex,\vertexindexi}$ that is the traffic processing capacity 
of a VM (e.g., in terms of compute requirements) for the logical module 
$\vertex_{\attackindex,\vertexindexi}$, where 
$\vertex_{\attackindex,\vertexindexi}$ is node $\vertexindexi$ of graph 
$\graph^{\logical}_\attackindex$.

\mypara{Network footprint} We denote  the  network-level cost  of transferring 
the unit of traffic from ingress  $\edgepopindex$ to datacenter
$\datacenterindex$  by $\cost_{\edgepopindex,\datacenterindex}$;
e.g., this can represent  the path latency per byte of traffic. Similarly,
within a datacenter, the units of intra-rack  and inter-rack traffic  costs
are denoted by 
$\unitIntraRackCost$ and $\unitInterRackCost$, 
respectively (e.g., they may represent latency
such that $\unitIntraRackCost < \unitInterRackCost$).

\subsection{Problem statement}
\label{subsec:problem_statement_res}



Our resource management problem is to translate  the annotated
graph into a  {\em physical graph} (see Figure~\ref{fig:resource}); i.e.,  each 
node $\vertexindexi$ of the annotated graph $\graph^{\logical}_\attackindex$ 
will be realized by one or more VMs each of which implement the 
logical module $\vertex_{\attackindex,\vertexindexi}$. 

\begin{figure}[t]
\begin{center}
\subfloat[Annotated graph.]
{
 \includegraphics[width=70pt]{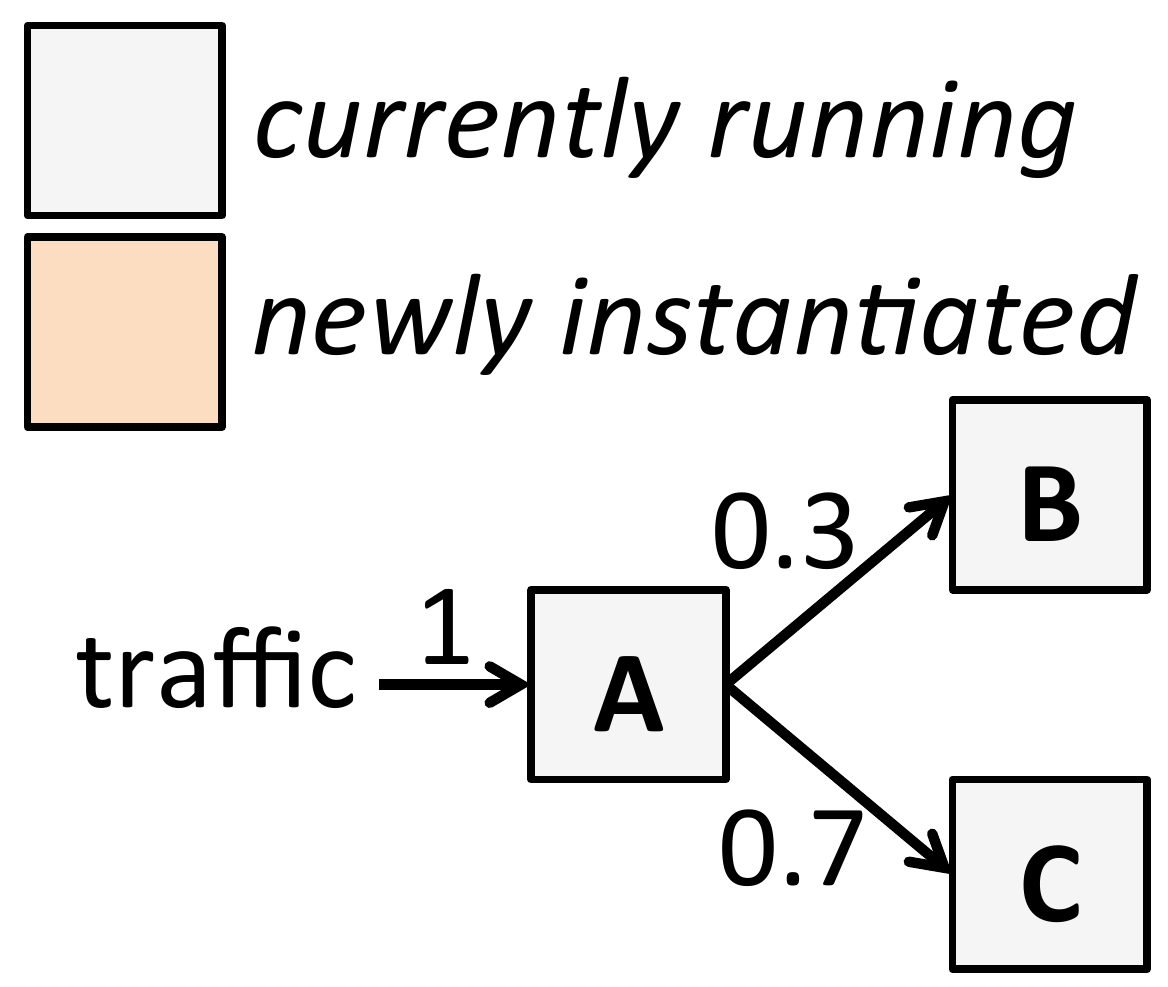}\hspace{0em}
\label{fig:scale_out_intuition_base}
}
\subfloat[Monolithic.]
{
 \includegraphics[width=80pt]{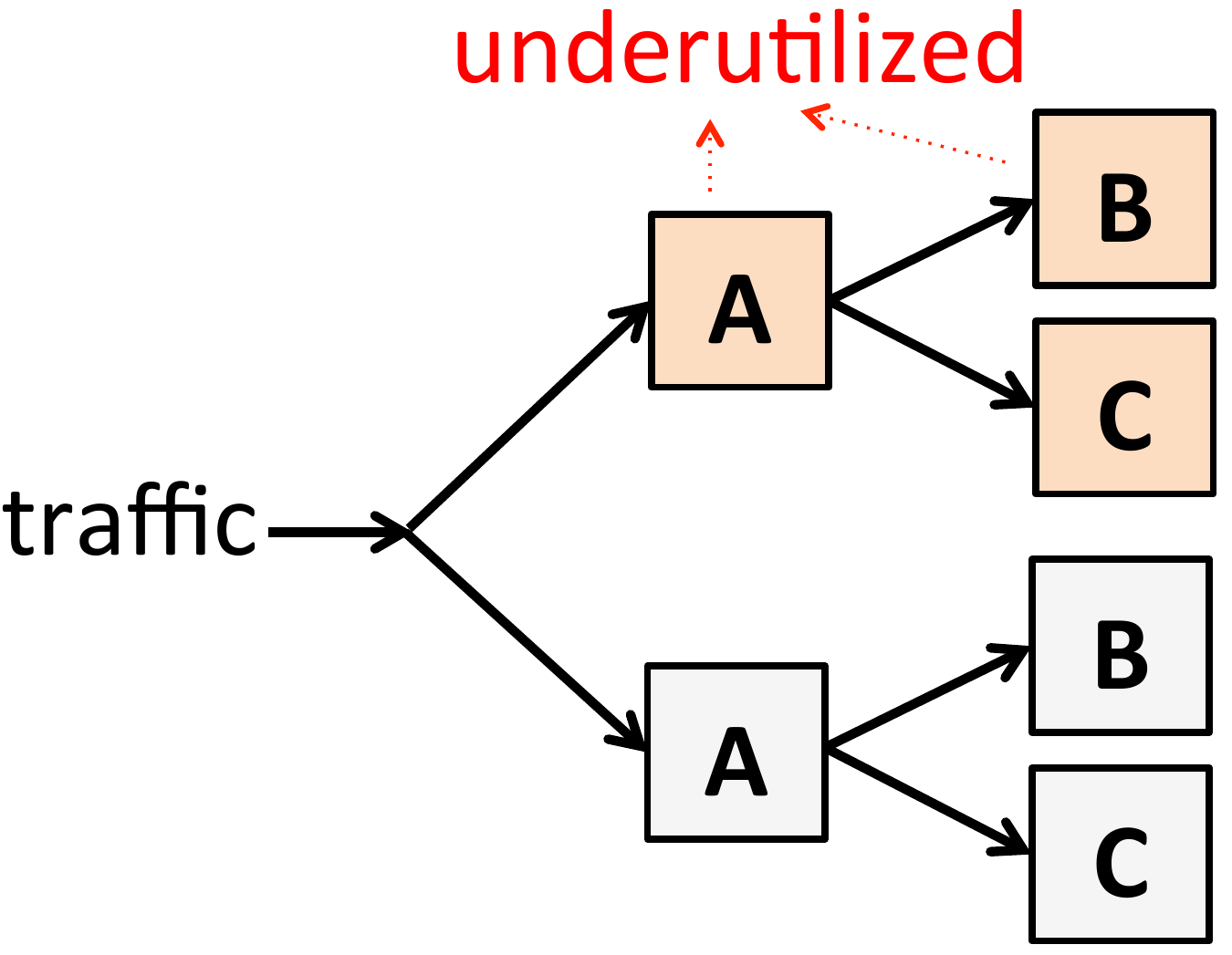}\hspace{0em}
\label{fig:scale_out_intuition_monolithic}
}
\subfloat[Fine-grained.]
{
 \includegraphics[width=70pt]{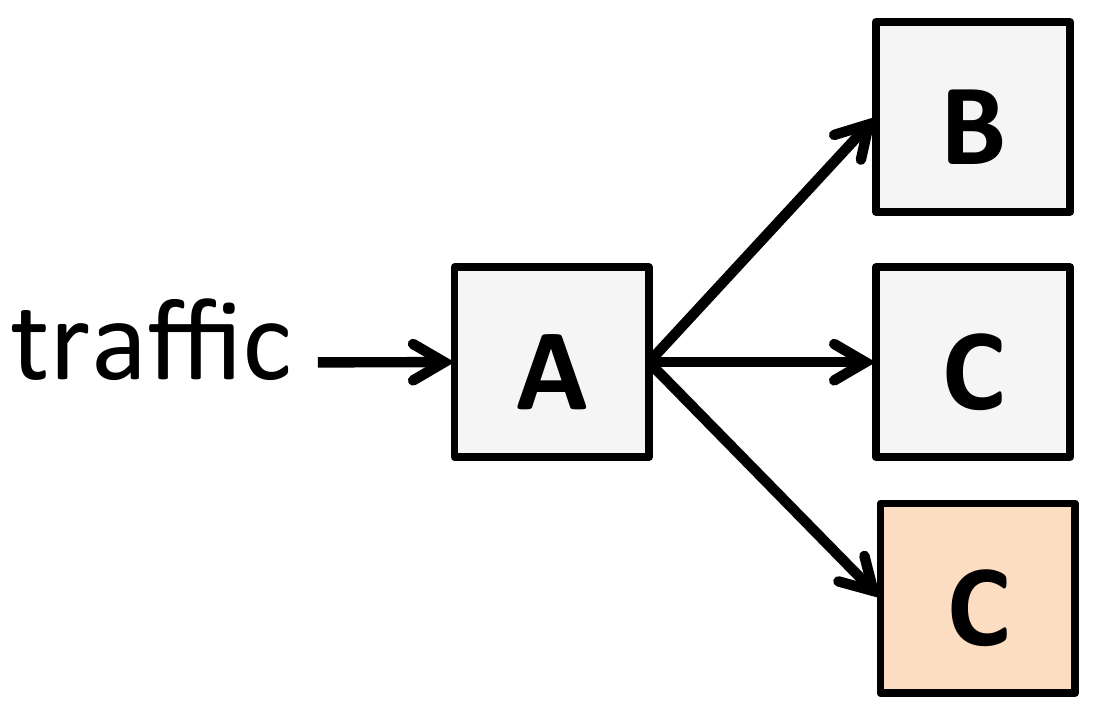}
\label{fig:scale_out_intuition_decoupled}
}
\end{center}
 \vspace{-0.3cm}
\tightcaption{An illustration of fine-grained elastic scaling when module
C becomes the  bottleneck.}
\label{scale_out_intuition}
 \vspace{-0.3cm}
 \end{figure}

\mypara{Fine-grained scaling} To generate physical graphs given annotated 
graphs in a resource-efficient manner, we adopt a {\em fine-grained scaling} 
approach, where each logical module is scaled independently.
We illustrate this idea in  Figure~\ref{scale_out_intuition}.
Figure~\ref{fig:scale_out_intuition_base} shows an annotated graph with three
logical modules A, B, and C, receiving  different amounts of traffic and consuming  
different  amounts of compute resources.  Once implemented as a physical graph,
suppose module C becomes the
bottleneck due to its processing capacity and input traffic volume.  Using a 
monolithic approach (e.g., running A, B, and C within a
single  VM), we will need to scale the entire graph as shown in 
Figure~\ref{fig:scale_out_intuition_monolithic}. Instead, we decouple  the 
modules to enable scaling out individual VMs; this yields higher resource efficiency 
as shown in Figure~\ref{fig:scale_out_intuition_decoupled}.




\shepherd{A-3 and C-3: we have dropped a main part of the notation here and only
maintained notations for the parameters that are important for the clarity of the section.}

\mypara{Goals} Our objective here is to  (a) instantiate the VMs across the
compute servers throughout the ISP, and (b) distribute the
processing  load across these servers to minimize the expected latency for
legitimate traffic. Further, we want to achieve (a) and (b) while minimizing 
the footprint of suspicious traffic.\footnote{While it is possible to 
explicitly minimize network congestion~\cite{snips_iciss}, minimizing suspicious 
traffic footprint naturally helps reduce network congestion as well.}

To this end, we need to assign values to two key sets of decision 
variables: (1)
 the fraction
of traffic  $\Traffic_{\edgepopindex,\attackindex}$ to send to each datacenter
$\datacenter_\datacenterindex$ (denoted by 
$\fraction_{\edgepopindex,\attackindex,\datacenterindex}$), and (2)
 the number of VMs of type  $\vertex_{\attackindex,\vertexindexi}$ to
run on server $\aserver$ of datacenter $\datacenter_\datacenterindex$. Naturally, these
decisions must  respect the  datacenters'  bandwidth and compute constraints.

 Theoretically, we can formulate this resource management problem as a
constrained optimization via an Integer Linear Program (ILP). For completeness,
we describe the full ILP in Appendix~\ref{subsec:ilp}. Solving the ILP
formulation gives an optimal solution to the resource management problem.
However, if the ILP-based solution is incorporated into \Name, an adversary can
easily overwhelm the system. This is because the ILP approach takes several hours
(see Table~\ref{tab:res_manager_timeandopt}). By the time it computes a
solution, the adversary may have radically changed the attack mix.

\subsection{Hierarchical decomposition}
\label{sec:dist}

 To solve the resource management problem,  we decompose the optimization 
 problem into two subproblems: (1) the  \Name global controller
solves a  \emph{Datacenter Selection Problem (DSP)}  to choose datacenters
responsible for processing suspicious traffic, and  (2) given the solution to the DSP, each  
local controller solves a \emph{Server Selection Problem (SSP)} to assign 
servers inside each selected datacenter to run the required VMs.
This decomposition is naturally  scalable  as the individual SSP problems can
be solved independently by datacenter controllers. Next, we describe  practical
greedy heuristics for the DSP and SSP problems that  yield
close-to-optimal solutions  (see Table~\ref{tab:res_manager_timeandopt}).

\mypara{Datacenter selection problem (DSP)} We design a greedy algorithm 
to solve DSP with the goal of reducing ISP-wide suspicious traffic footprint.
To this end, the algorithm first sorts suspicious traffic volumes  
(i.e., $\Traffic_{\edgepopindex,\attackindex}$ values) in a decreasing order. 
Then, for each  suspicious traffic volume $\Traffic_{\edgepopindex,\attackindex}$ 
from the sorted list, the algorithm tries to assign the traffic volume to the datacenter with the 
least cost based on $\cost_{\edgepopindex,\datacenterindex}$ values. The 
algorithm has two outputs: (1) $\fraction_{\edgepopindex,\attackindex,\datacenterindex}$ 
values denoting what fraction of suspicious traffic from each ingress should be
steered to each datacenter (as we will see in
\Section\ref{sec:orchestration_layer}, these values will be used by network
orchestration to steer traffic correspondingly), (2) the physical graph
corresponding to attack type $\attackindex$ to be deployed by datacenter
$\datacenterindex$.  For completeness, we show the pseudocode for the 
DSP algorithm in Figure~\ref{fig:dsp} in Appendix~\ref{subsec:dsp_ssp}.



\mypara{Server selection problem (SSP)}  Intuitively, the SSP algorithm attempts
to preserve traffic locality by instantiating nodes  adjacent in the physical  graph as
close as possible within the datacenter.  Specifically, given the physical
graph, the SSP algorithm  greedily  tries to assign \nodes with higher capacities
(based on $\power_{\attackindex,\vertexindexi}$ values) along with its
predecessors to the same server, or the same rack.  For completeness we show
the pseudocode for the SSP algorithm in Figure~\ref{fig:ssp} in 
Appendix~\ref{subsec:dsp_ssp}.




 
\section{Network Orchestration}
\label{sec:orchestration_layer}

Given the outputs of the resource manager module (i.e., assignment of
datacenters to incoming suspicious traffic and assignment of servers to defense
VMs), the role of the network orchestration module is to configure the network to 
implement these decisions.  This includes setting up forwarding rules in the ISP 
backbone and inside the datacenters. The  main  requirement is scalability 
in the presence of attack traffic.  In this section, we   present our {\em tag-based} 
and {\em proactive} forwarding approach to address the limitations of  the 
per-flow and reactive SDN approach.

 

\subsection{High-level idea}
\label{subsec:orchestration_strawman}

As discussed earlier in \Section\ref{subsec:challenges}, the canonical SDN 
view of setting up switch forwarding rules in a per-flow and reactive manner 
is not suitable in the presence of \ddos attacks. Furthermore, there are practical 
scalability and deployability concerns  with using SDN in ISP 
backbones~\cite{sdn_sequel, fabric}.  There are two main ideas in our approach 
to address these limitations:

\begin{packeditemize}

\item  Following the hierarchical decomposition in resource management, we 
also decompose the network orchestration problem into two-sub-problems: 
(1) Wide-area routing to get traffic to datacenters, and (2) Intra-datacenter routing 
to get traffic to 
the right VM instances. This decomposition allows us to use different 
network-layer  techniques; e.g., SDN is more suitable inside  the datacenter  
while traditional MPLS-style routing is better suited for wide-area routing. 

\item Instead of the controller reacting to each flow arrival, we {\em
proactively} install forwarding  rules before traffic arrives. Since we do not
know the specific IP-level suspicious flows that will arrive in the future, 
we use logical {\em tag-based}  forwarding rules with per-VM tags 
instead of per-flow rules.
 
\end{packeditemize}


\subsection{Wide-area orchestration} 

The \Name global controller sets up forwarding rules on backbone routers 
so that traffic detected as suspicious is steered from edge PoPs to datacenters 
according to the resource management decisions specified by the
$\fraction_{\edgepopindex,\attackindex,\datacenterindex}$ values (see 
\Section\ref{sec:dist}).\footnote{We assume the ISP uses legacy mechanisms 
for forwarding non-attack traffic and traffic to non-\Name customers, so 
these are not the focus of our work.} 

 To avoid a forklift upgrade of the ISP backbone and enable
an immediate  adoption of \Name, we  use traditional tunneling
mechanisms in the backbone (e.g., MPLS or IP tunneling). 
\comment{On the ISP backbone, we need SDN support only at the ISP edge 
routers---this is to handle stateful processing as we will discuss at the end 
of this section) while all other  backbone switches do not need an
upgrade to support SDN.
\vyas{UNCLEAR: The only SDN-like 
 capability we need in the backbone is at the ISP edge routers (to 
handle stateful processing as we will discuss at the end of this section),
which is a relatively easy upgrade compared to upgrading all backbone 
switches.}\seyed{check the fix.}
}
We proactively set up static tunnels from each edge PoP 
to each datacenter. Once the global 
controller has solved the DSP problem, the controller configures 
backbone routers to split the traffic according to the 
$\fraction_{\edgepopindex,\attackindex,\datacenterindex}$
values. While our design is not tied to any specific 
tunneling scheme, the widespread use of MPLS and IP tunneling 
make them natural candidates~\cite{google_sdn}. 

\comment{
Figure~\ref{fig:tunnel} shows an example of how \Name performs inter 
datacenter routing.

\begin{figure}[h]
\centering
\includegraphics[width=250pt]{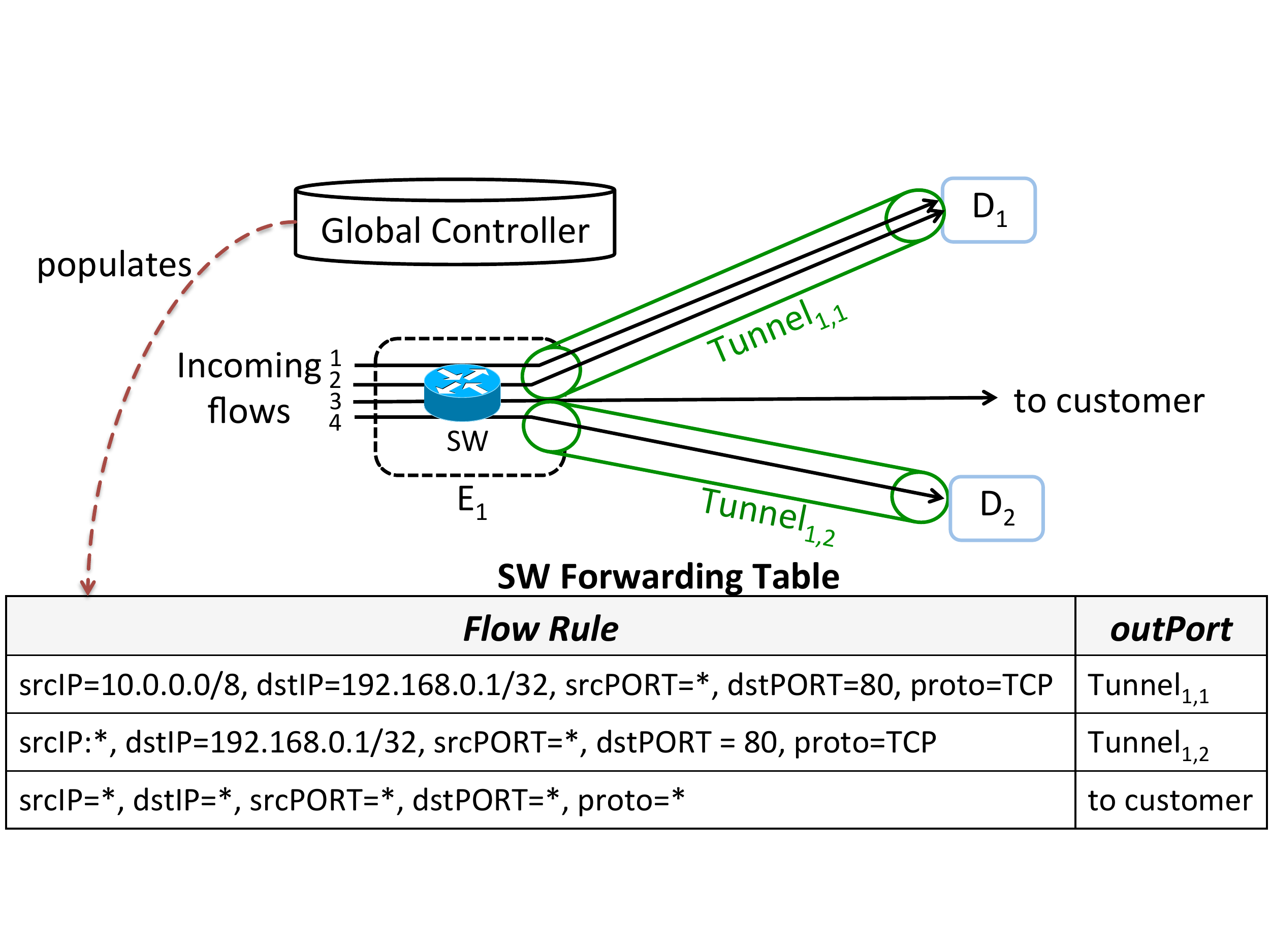}
\caption{Tunnel-based global routing in \Name. (Rules in upper rows are 
assumed to have a higher match priority.)}
\label{fig:tunnel}
\end{figure}
}

\subsection{Intra-datacenter orchestration}
Inside each datacenter, the  traffic needs to be steered through the intended
sequence of VMs.  There are two main considerations here: 

\begin{packedenumerate}

\item The next  VM a packet needs to be sent to 
depends on the \emph{context} 
of the current VM. For example, the node \emph{check UDP count 
of src} in the graph shown in Figure~\ref{example_graph} may send traffic 
to either \emph{forward to customer} or \emph{log} depending on its 
analysis outcome.

\item With  elastic scaling, we may instantiate several  physical VMs
for each logical node depending on the demand.  Conceptually, we need a ``load
balancer'' at every level of our annotated graph to distribute traffic across
different VM instances of a given logical node. 

\end{packedenumerate}

Note that we can trivially  address both requirements using  a per-flow and 
reactive solution. Specifically,  a local controller can track a packet as it 
traverses the physical graph,  obtain the
relevant context information from  each VM, and determine the next VM to
route the traffic to. However, this approach is clearly not scalable and 
 can introduce avenues for new attacks. The challenge here
 is to meet these requirements without incurring the overhead of this
 per-flow and reactive approach.

\begin{figure}[t]
\centering
\includegraphics[width=210pt]{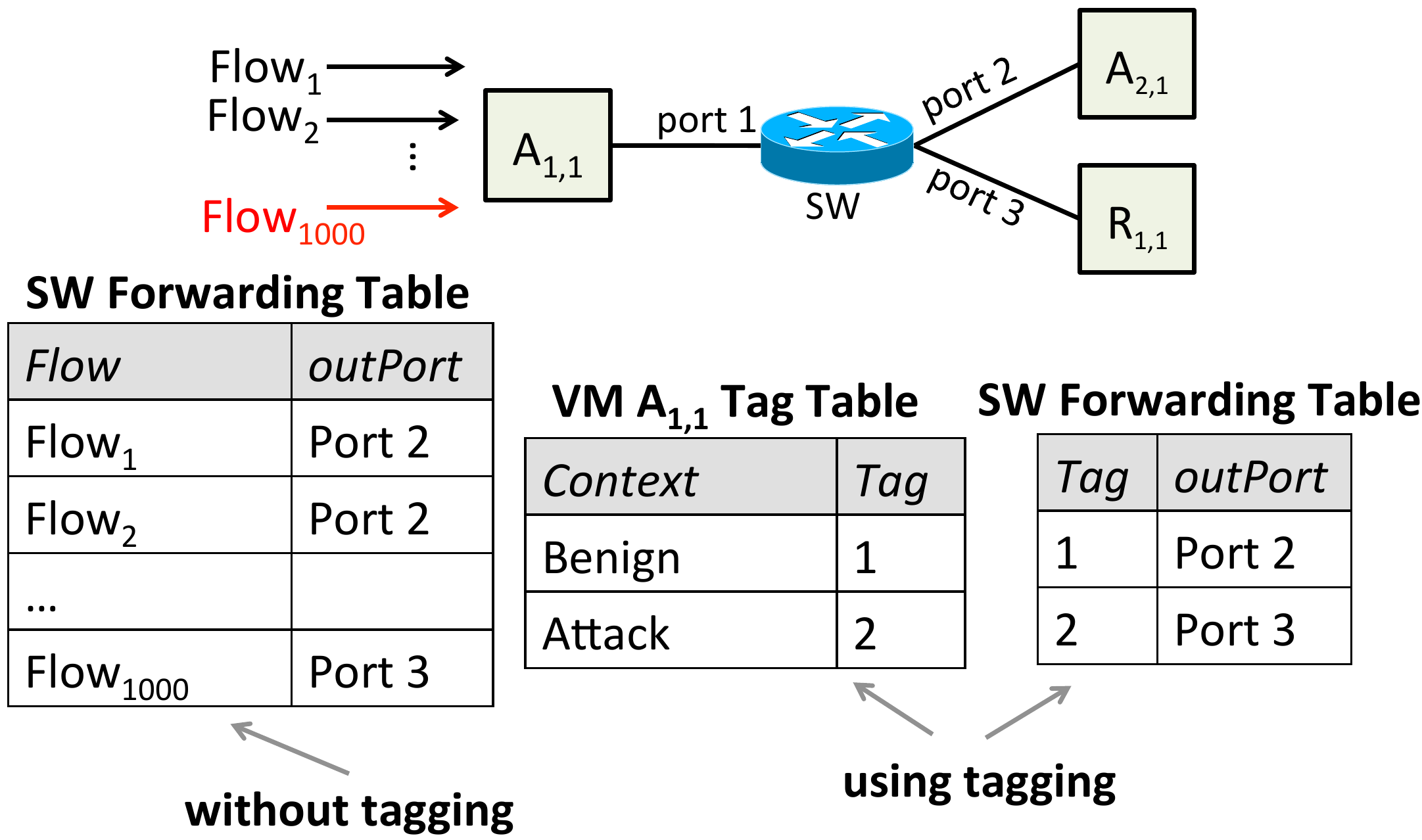}
\vspace{-0.2cm}
\tightcaption{Context-dependent forwarding using tags.}
\vspace{-0.3cm}
\label{fig:context}
\end{figure}

\mypara{Encoding processing context} Instead of having the controller
track the context, our high-level idea is to encode the necessary context as  \emph{tags} inside
packet headers~\cite{flowtags_nsdilong}. 
  Consider the example shown
in Figure~\ref{fig:context} composed  of VMs $A_{1,1}$, $A_{2,1}$,
and $R_{1,1}$. $A_{1,1}$ encodes the processing context of outgoing traffic as
tag values embedded in its outgoing packets (i.e., tag values 1 and 2 
denote benign and attack traffic, respectively). The switch then uses this tag 
value to forward each packet to the correct next VM.


Tag-based forwarding addresses the control channel bottleneck and
switch rule explosion.  First, the tag generation and tag-based
forwarding behavior of each VM and switch is configured proactively once
the local controller has solved the SSP.  We 
proactively assign a tag for each VM and populate forwarding rules before flows
arrive; e.g., in Figure~\ref{fig:context}, the tag table of $A_{1,1}$ and the
forwarding table of the router have been already populated as shown.
 Second, this reduces router forwarding rules as illustrated in
Figure~\ref{fig:context}. Without tagging, there will be one rule for each of
the 1000  flows.  Using tag-based forwarding, we achieve the same forwarding
behavior using only two forwarding rules.

\begin{figure}[t]
\centering
\subfloat[A naive \mux design.]
{
\vspace{-0.6cm}
\includegraphics[width=57pt]{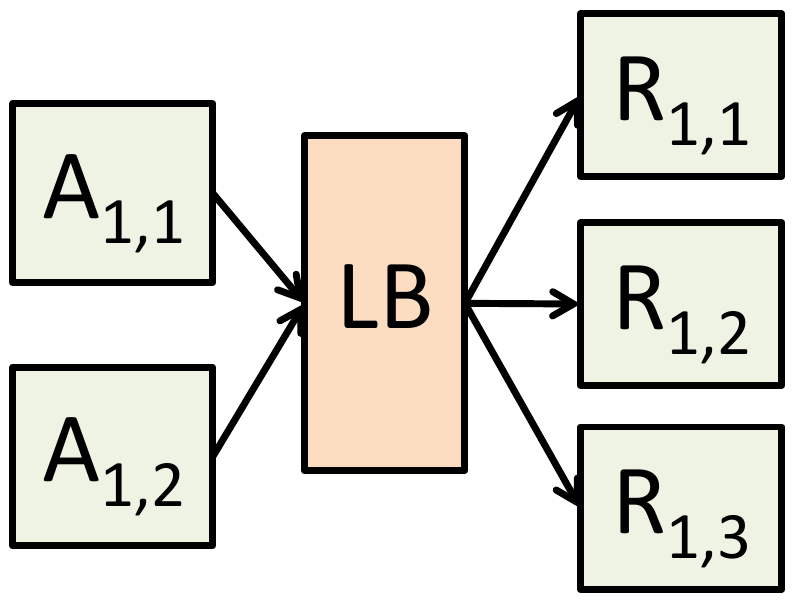}
\label{fig:naive_mux}
}
\subfloat[A distributed \mux design.]
{
\vspace{-0.6cm}
\includegraphics[width=165pt]{figs/new_tag_based_forwarding_2.pdf}
\label{fig:dist_mux}
}
  \tightcaption{Different \mux design points.}
\vspace{-0.3cm}
\label{fig:mux_design}
\end{figure}

\mypara{Scale-out load balancing}  One could interconnect VMs of the same
physical graph as shown in
Figure~\ref{fig:naive_mux} using a dedicated load balancer (\mux). 
However, such a \mux may itself become a bottleneck, 
as it is on the path of  every packet from any VM in the set
$\{A_{1,1},A_{1,2}\}$ to any VM in the set $\{R_{1,1},R_{1,2},,R_{1,3}\}$. 
To circumvent this problem,   
we implement the  \muxing strategy \emph{inside each VM} 
so that the \mux capability scales proportional to the current number of 
VMs. 
 Consider the example shown in Figure~\ref{fig:dist_mux} 
 where  due to an increase in attack traffic volume we have added one 
more  VM of type $A_1$ (denoted by $A_{1,2}$) and one more VM of 
type  $R_1$ (denoted by $R_{1,2}$). 
To load balance traffic between the two VMs of type $R_1$, the \mux  of $A_1$ 
VMs (shown as $LB_{1,1}$ and $LB_{1,2}$ in the figure) pick a tag 
value from a \emph{tag pool} (shown by \{2,3\} in the figure) based on the 
processing context of the outgoing packet and the intended load balancing 
 scheme (e.g., uniformly at random to distribute load equally). Note that this 
 tag pool is pre-populated by the local controller (given the defense library 
 and the output of the resource manager module).  This scheme, thus, satisfies 
 the load balancing requirement in a scalable manner. 

\comment
{\begin{figure}[h]
\centering
\includegraphics[width=220pt]{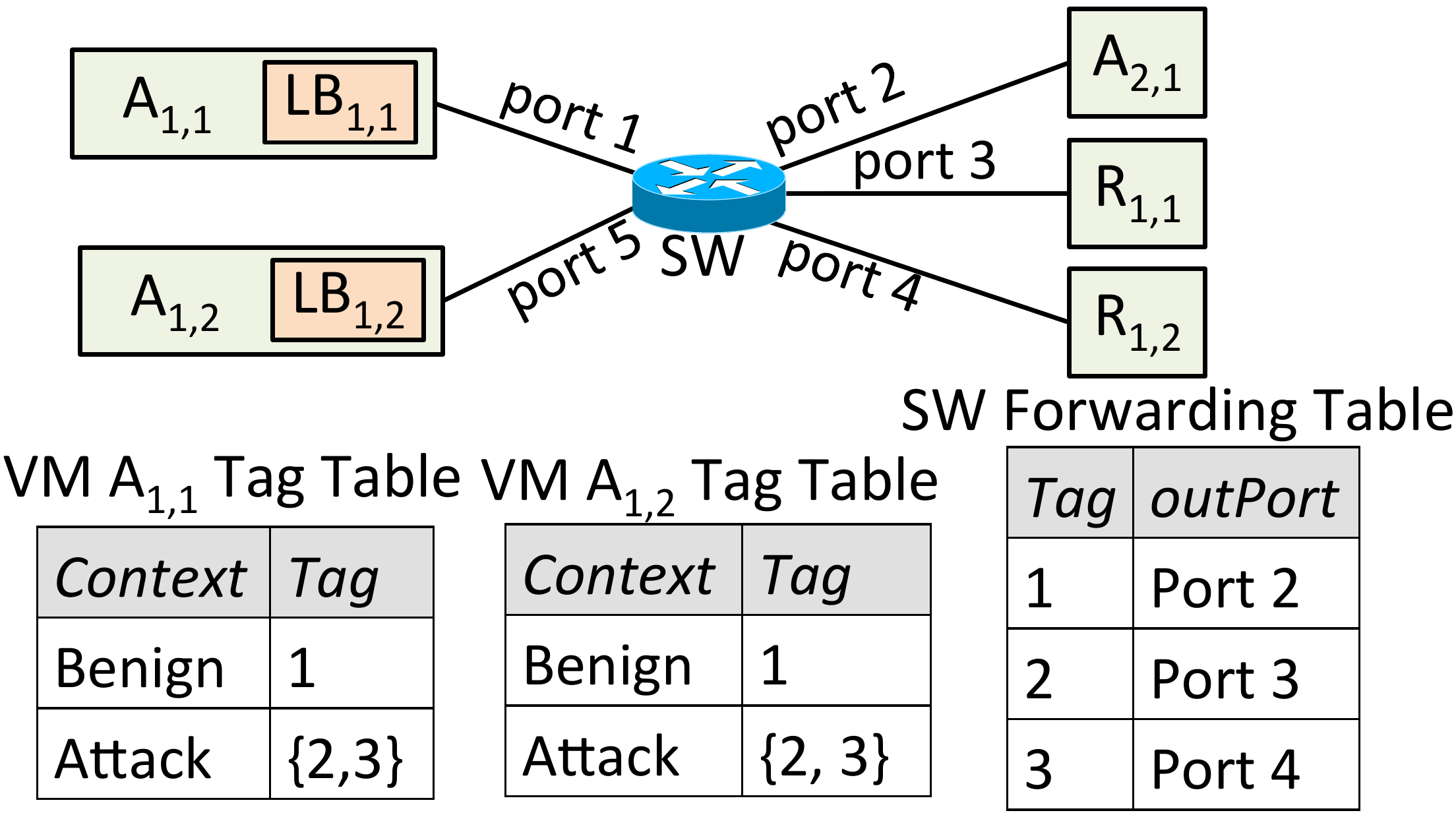}
\caption{Load balancing between $R_{1,1}$ and $R_{1,2}$ using tag pools.}
\label{fig:loadbalancing}
\end{figure}
}



\mypara{Other issues} There are two remaining practical issues:
\begin{packeditemize}

\item {\em Number of tag bits:}  We give a simple upper bound on the required number of
bits to encode tags.  First, to support context-dependent forwarding out of a VM with
$k$ relevant contexts, we need $k$ distinct tag values. Second. to support load
balancing among $l$ VMs of the same logical type, each VM needs to be populated with a
tag pool including $l$ tags.  Thus, at each VM we need at most $k\times l$
distinct tag values. Therefore, an upper bound on the total number of unique 
tag values  is  $k_{max} \times l_{max} \times \sum 
\limits_{\attackindex} \mid \vertexes_\attackindex^{\logical} \mid$, where
$ k_{max}$ and $l_{max}$ are the maximum number of contexts and VMs
of the same type in a graph, and $\vertexes_\attackindex^{\logical}$ is the set
of vertices of annotated graph for attack type \attackindex. To make this concrete,
across the evaluation experiments \Section\ref{sec:evaluation}, the maximum
value required tags was 800, that can be encoded in $\ceil{log_2 (800)}=10$
bits. In
practice, this tag space requirement of \Name can be easily satisfied given that
datacenter grade  networking platforms already have extensible  header
fields~\cite{vmware}.

\item {\em Bidirectional processing:}  Some logical modules may have bidirectional 
semantics. For example, in case of  a DNS amplification attack, request and 
response traffic must be processed by the same VM. (In other cases, such as 
the UDP flood attack, bidirectionality is not required.). To enforce bidirectionality, 
ISP edge switches use tag values of outgoing traffic so that when the 
corresponding incoming traffic comes back, edge switches sends it to the 
datacenter within which the VM that processed the outgoing traffic is located. 
Within the datacenter, using this tag value, the traffic is steered to the VM.

\end{packeditemize}

\section{Strategy Layer}
\label{sec:policy_layer}

As we saw in \Section\ref{sec:resource_layer}, a key input to the 
resource manager module is the set of $\Traffic_{\edgepopindex,\attackindex}$ 
values, which represents the volume of  
suspicious traffic of each attack type $\attackindex$ arriving at  each edge PoP 
\edgepopindex. This means we need to  estimate the future attack mix 
based on observed measurements of the 
network and  then instantiate the  required defenses.  We begin by describing 
an adversary that intends to thwart a \Name-like system.  Then, we discuss 
limitations of  strawman solutions  before describing our {\em online adaptation} 
mechanism.


\mypara{Interaction model} We model the interaction between the ISP 
running \Name and the \adversary as a
repeated interaction over several {\em epochs}. The ISP's ``move'' is one epoch
behind the \adversary; i.e., it takes \Name an epoch to react to a new attack
scenario due to implementation delays  in \Name operations. 
 The epoch duration is simply the sum of the time to detect the attack,
run the resource manager, and execute the network orchestration logic.
While  we can engineer the system to minimize this lag, there will 
still be non-zero delays in practice and thus we need an adaptation strategy.

\mypara{Objectives} Given this interaction model, the ISP has to pre-allocate VMs
and hardware resources for a specific attack mix.
An intelligent and dynamic adversary can change its 
 attack mix to meet two goals: 

\begin{packeditemize}

\item[{\bf G1}] {\em Increase hardware resource consumption:}  The adversary can cause 
ISP to overprovision defense VMs. This may impact the ISP's ability to accommodate
other attack types or reduce  profits from other services that could have
 used the infrastructure. 

\item[{\bf G2}] {\em Succeed in delivering attack traffic:} If the ISP's  detection and 
 estimation logic is sub-optimal and does not have the required defenses installed, then the adversary can 
  maximize the volume of attack traffic delivered to the target. 


\end{packeditemize}

The adversary's goal is to maximize these objectives, while the ISPs goal is 
 to minimize these to the extent possible. 
 One could also consider a third objective of collateral damage on legitimate traffic;
e.g., introduce needless delays.  We do not discuss this dimension  because our
optimization algorithm from \Section\ref{sec:resource_layer} will naturally
push  the defense as close  to the ISP edge (i.e., traffic ingress points) as 
possible to minimize  the impact on legitimate traffic.

\mypara{Threat model} We consider an  \adversary with a fixed budget 
in terms of the total volume
of attack traffic it can launch at any given time.
 Note that the adversary  can apportion this budget across the {\em
types} of attacks and the {\em ingress} locations from which the attacks are
launched.  Formally,  we have $\sum\limits_{\edgepopindex}^{}
\sum\limits_{\attackindex}^{} \Traffic_{\edgepopindex,\attackindex} \leq
\attackBudget$, but there are  no constraints on the specific
$\Traffic_{\edgepopindex,\attackindex}$ values.

\comment
{
Therefore, we define the cost imposed to the ISP by the dynamic 
adversary's by accounting for the above three cost components as 
follows:\\

\noindent $\Cost_{adv.} = \coeffVMs \times |\VMs| + 
\coeffAttack \times |\UnstoppedAttack| +
(1- \coeffVMs - \coeffAttack) \times |\SuspiciousLegit|$\\

\noindent where $|\VMs|$ is the total of VMs run by \Name,
$|\SuspiciousLegit|$ is the total volume of the legitimate component
of suspicious traffic, and $|\UnstoppedAttack|$ is the total volume
of attack traffic that reaches the customer. $\coeffVMs$ and $\coeffAttack$
are ISP-dependent normalization factors that show the relative 
importance of each term in determining the total cost.
}


\mypara{Limitations of strawman solutions}
  For simplicity, let us consider a single
ingress point. Let us consider a  strawman solution
called  {\em PrevEpoch} where we measure the attack observed in the previous epoch
and use it as the estimate for the next epoch. 
Unfortunately, this can have serious issues 
w.r.t.\ goals   {\bf G1} and {\bf G2}.
To see why, consider a simple scenario where we have two attack types
with a budget of 30 units and three epochs with the attack volumes as follows:
T1: A1= 10, A2=0; T2: A1=20, A2=0; T3: A1=0; A2=30. 
Now consider the {\em PrevEpoch} strategy starting at the 0,0 configuration.  It
has a   total wastage of 0,0,20 units and  a total evasion of 10,10,30
units because it has overfit to the previous measurement.   
We can also consider other strategies; e.g., a  {\em Uniform} strategy that
provisions 15 units each for A1 and A2 or extensions of these to {\em
overprovision} where we multiply the  number of VMs given by the resource
manager in the last epoch by a fixed value $\overprovisionFactor > 1$. However,
these suffer from the same problems and are not competitive. 

\mypara{Online adaptation}
 Our metric
of success here is to have {\em low regret} measured with
respect to the best static solution computed in hindsight~\cite{kalai}. 
 Note that in
general, it is not possible to be competitive w.r.t.\ the best dynamic solution
since that presumes oracle knowledge of the adversary, which is not practical.



Intuitively, if we have a non-adaptive adversary, using the observed empirical
average is the best  possible static hindsight estimation strategy; i.e.,
$\Traffic^*_{\edgepopindex,\attackindex} = \frac{\sum_{\epoch}
\Traffic_{\edgepopindex,\attackindex,\epoch}}{|\epoch|}$ 
 would be the optimal solution ($|\epoch|$ denotes the total number of epochs). However,
an attacker who knows that we are using this strategy can game the system by 
changing the attack mix. To address this, we use a follow the perturbed leader
(FPL) strategy~\cite{kalai} where our estimation   uses a combination of the past observed
behavior of the adversary and a randomized component. Intuitively, the random 
component makes it impossible for the attacker to predict the ISP's estimates. 
This is a well-known approach in online algorithms to minimize the 
regret~\cite{kalai}.  
Specifically,  the traffic estimates  for the $\mathit{next}$ epoch
$t+1$, denoted by $\widehat{\Traffic_{\edgepopindex,\attackindex,\epoch+1}}$
values, are calculated
based on the average of the past values plus a random component:
$\widehat{\Traffic_{\edgepopindex,\attackindex,\epoch+1}} =  \frac{\sum_{\epoch'=1}^{\epoch} \Traffic_{\edgepopindex,\attackindex,\epoch'}}{|\epoch|}  + \mathit{randperturb}$.


Here, $\Traffic_{\edgepopindex,\attackindex,\epoch'}$ is the empirically observed value of the attack 
traffic and  $\mathit{randperturb}$ is a random value drawn uniformly from 
$[0,\frac{2 \times \attackBudget}{nextEpoch \times |\edgepop| \times |\maxAttack|}]$.
(This is assuming a total defense of budget of  $2 \times \attackBudget$.)
 It can be shown that this is indeed a provably good regret minimization 
 strategy~\cite{kalai}; we do not show the proof for brevity.

\section{Implementation}
\label{sec:implementation}




In this section,  we briefly describe how we implemented the key functions 
described in the previous sections. We have made the source code 
available~\cite{bohatei_code}.

\subsection{\ddos defense \modules}

The design of the \Name strategy layer is inspired by  the prior modular 
efforts in Click~\cite{click} and Bro~\cite{bro}.  This modularity
has two advantages. First, it allows us to adopt best of breed solutions
and compose them for different attacks. Second, it enables more fine-grained 
scaling.  At a high level, there are two types 
of logical building blocks in our defense library:

\begin{packedenumerate}

\item \emph{Analysis (A)}:   
Each analysis module processes a suspicious flow and  determines 
appropriate action  (e.g., more analysis or specific response).
 It receives a  packet  and outputs a tagged packet,
 and the tags are used to steer traffic to subsequent analysis and response 
\module instances as discussed earlier.


\item \emph {Response (R)}:   The input to an R module
is a tagged packet from some A module. Typical responses include
forward to customer (for benign traffic), log, drop, and rate
limit.  Response functions will depend on the type of  attack; e.g., sending
RST packets in case of a TCP SYN attack. 

\end{packedenumerate}

Next, we describe defenses we have  implemented for different \ddos attacks.
Our goal here is to illustrate the flexibility \Name provides in dealing with a 
diverse set of known attacks rather than develop new defenses.

\begin{figure}[t]
 \includegraphics[width=230pt]{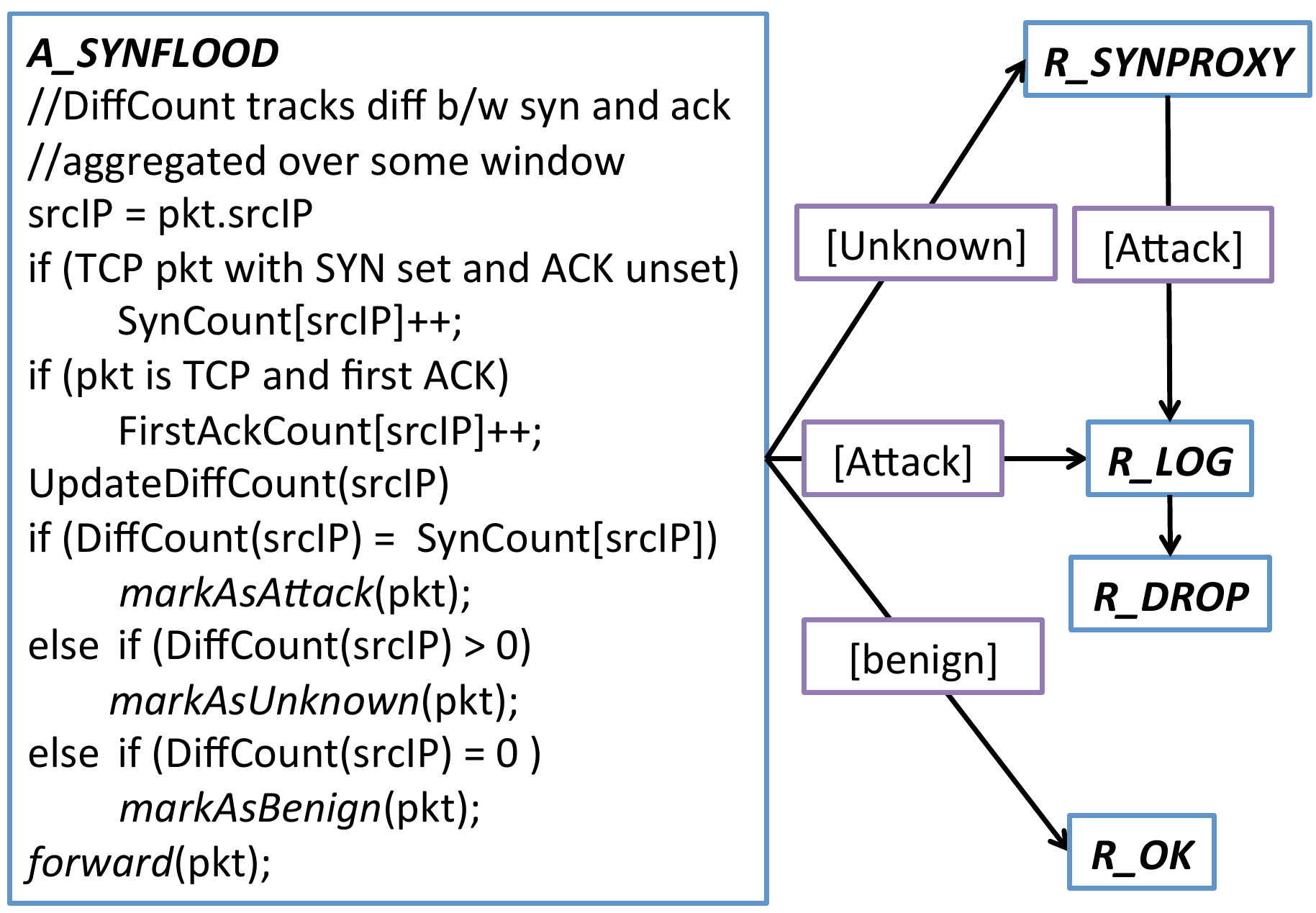}
 \vspace{-0.5cm}
\tightcaption{SYN Flood defense strategy graph.}

\label{fig:new_syn_graph}
\end{figure}
\begin{figure}[t]
 \includegraphics[width=240pt]{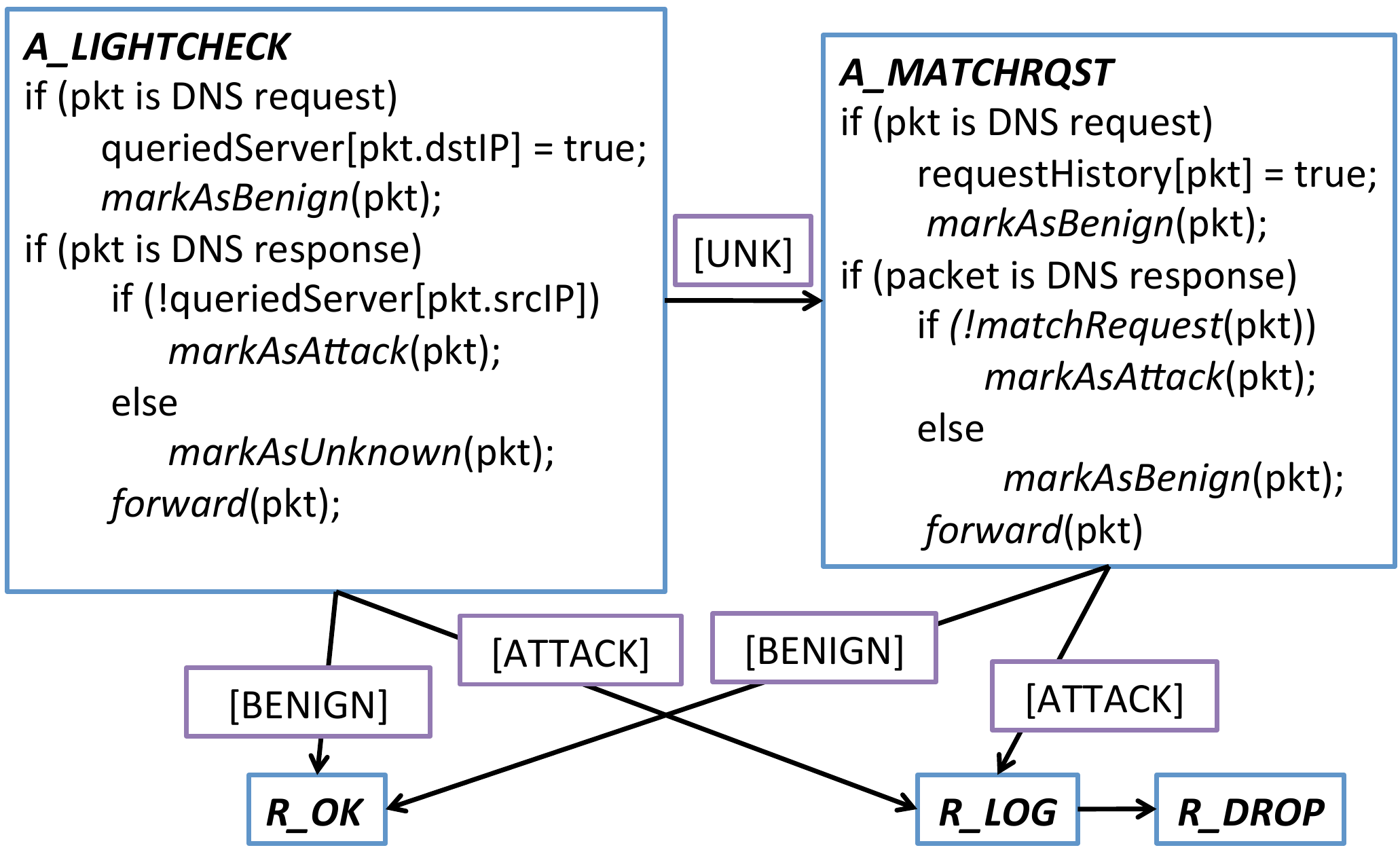}
 \vspace{-0.5cm}
\tightcaption{DNS amplification defense strategy graph.}
\label{fig:new_dns_graph}
\vspace{-0.2cm}
\end{figure}

\begin{packedenumerate}

\item {\sc \bf SYN flood} (Figure~\ref{fig:new_syn_graph}):  We track the
number of open TCP sessions for each source IP; if a source IP has no
asymmetry between SYNs and ACKs, then mark its packets  as benign. If a 
source IP never completes a connection, then we can mark its future packets 
as known attack packets.  If we see a gray area where  the source IP has 
completed some connections but not others, in which case we use a 
SYN-Proxy defense
(e.g.,~\cite{synproxy_redhat,synproxy_paper}).


\item {\sc \bf DNS amplification} (Figure~\ref{fig:new_dns_graph}): We check if
the DNS server  has been queried by some customer IP. This example highlights
another advantage---we can decouple fast (e.g., the header-based
$\mathit{A\_LIGHTCHECK}$  module) and slow path analyses  (e.g., the second
A module needs to look into payloads). The  responses are quite simple
and implement  logging,  dropping,  or basic forwarding to the destination.   
We do not show the code  for brevity.

\item {\sc \bf UDP flood:}  The
analysis node $\mathit{A\_UDP}$ identifies source IPs that send an anomalously 
higher number of UDP packets and  uses this
 to categorize each packet as either attack or benign.   The
 function  $\mathit{forward}$ will direct the packet to the next  node in the
 defense strategy; i.e.,  $\mathit{R\_OK}$ if benign,  or $\mathit{R\_LOG}$ if
attack.  


\item {\sc \bf Elephant flow:} Here, the attacker launches legitimate but very
large flows.   The A module  detects abnormally large flows  and flags them 
as attack flows. The response  is to randomly drop packets from these large 
flows (not shown). 

\end{packedenumerate}

 \begin{table}[t]
  \begin{center}
  \begin{footnotesize}
  \begin{tabular}{p{1cm}|p{3cm}|p{3cm}}
         Attack type     & Analysis 		& Response\\ \hline

          UDP flood  &  A\_UDP using Snort (inline mode)	& R\_LOG using iptables and R\_RATELIMIT using tc library  \\ \hline

          DNS amp.  & both LIGHTCHECK and MATCHRQST  using netfilter library, iptables, custom code & R\_LOG and R\_DROP using iptables  \\ \hline 

          SYN flood	& A\_SYNFLOOD using Bro & R\_SYNPROXY using PF firewall, R\_LOG and R\_DROP using iptables \\ \hline 

          Elephant flow 	& A\_ELEPHANT using  netfilter library, iptables, custom code & R\_DROP using iptables 
     \end{tabular}
  \end{footnotesize}
  \end{center}
  \vspace{-0.4cm}
 \caption{Implementation of \Name modules.}
  \vspace{-0.4cm}
 \label{tab:implementation}
 \end{table}

\myparatight{Attack detection} We use simple time series anomaly detection using
{\tt nfdump}, a tool that provides NetFlow-like capabilities, and custom 
code~\cite{wavelet}.  The output of the detection module is sent to
the \Name global controller as a 3-tuple $\langle \mathit{Type, FlowSpec,
Volume} \rangle$, where $\mathit{Type}$ indicates the type of \ddos attack
(e.g., SYN flood,  DNS amplification), $\mathit{FlowSpec}$ provides a generic
description of the {\em flow space} of suspicious traffic (involving wildcards), 
and $\mathit{Volume}$ indicates the suspicious traffic volume based on 
the flow records. Note that this  $\mathit{FlowSpec}$ does not  pinpoint
specific attack flows; rather, it is a coarse-grained hint on characteristics of suspicious
traffic that need further processing through the defense graphs.


\subsection{SDN/NFV platform}

\myparatight{Control plane}  We use the {\tt OpenDayLight} network
control platform, as it has gained significant traction from key
industry players~\cite{odl}. We implemented the  \Name  global
and local control plane modules (i.e., strategy, resource management,
and network orchestration)  as separate  {\tt OpenDayLight} plugins.
\Name uses OpenFlow~\cite{openflow} for configuring switches; this is purely
for ease of prototyping, and it is easy to integrate other
network control APIs (e.g., YANG/NetCONF).


\myparatight{Data plane} Each physical \node is realized
using a VM running on KVM.  We  use open source tools (e.g., Snort, Bro)  to 
implement the different Analysis (A) and Response (R) \modules.  
Table~\ref{tab:implementation} summarizes the specific platforms we have 
used.   These tools are instrumented using FlowTags~\cite{flowtags_nsdilong} 
to add tags to outgoing packets to provide  contextual information.
 We used OpenvSwitch~\cite{openvswitch} to 
 emulate switches in both datacenters and ISP backbone.  The choice
of OpenvSwitch is for ease of prototyping on our testbed.   

\myparatight{Resource management algorithms}   We implement the
DSP and SSP algorithms using custom {\tt Go} code.




\section{Evaluation}
\label{sec:evaluation}


In this section, we  show that:

\begin{packedenumerate}

\item \Name is scalable and handles attacks of hundreds of Gbps 
 in large ISPs and  that our design decisions are crucial for 
 its scale and responsiveness (\Section\ref{subsec:benefits})

\item \Name enables a rapid ($\leq$ 1 minute) response for several canonical
\ddos attack scenarios (\Section\ref{subsec:validation})

\item \Name  can  successfully cope with several dynamic attack
strategies (\Section\ref{subsec:advesary_eval})

\end{packedenumerate}






\mypara{Setup and methodology} We use a combination of real testbed and
trace-driven evaluations to demonstrate the above benefits. Here we
briefly  describe  our testbed, topologies, and attack configurations:

\begin{packeditemize}

\item \emph{SDN Testbed:} Our  
testbed has 13 Dell R720 machines (20-core 2.8 GHz Xeon CPUs, 128GB RAM). Each machine runs 
 KVM on CentOS 6.5 (Linux kernel v2.6.32). On each machine, 
we assigned equal amount of resources to each VM: 1 vCPU (virtual CPU) 
and 512MB of memory. 
\comment{
Further, we  assigned a fixed number of physical 
CPU cores to each VM  to ensure  performance isolation across VMs without 
relying on the default  KVM resource  scheduling. }

\item \emph{Network topologies:} We emulate several router-level  ISP
topologies (6--196 nodes) from the Internet Topology
Zoo~\cite{topology_zoo}. We set the   bandwidth of each core link to be 100Gbps and 
 link latency  to be 10ms. The number of datacenters, which are located 
 randomly, is 5\% of the number of backbone switches with a capacity of 
 4,000 VMs per datacenter.


\item \emph{Benign traffic demands:}  We assume a gravity model of traffic
demands between ingress-egress switch pairs~\cite{r:05}. The total volume is scaled
linearly with the size of the network such that the average link load on the
topology backbone is 24Gbps with a maximum bottleneck link load of 55Gbps.  We
use {\tt iperf} and custom code to generate benign traffic.

\item \emph{Attack traffic:}  We implemented custom modules to generate attack
traffic: (1) {\sc \bf SYN flood} attack by sending only SYN packets with
spoofed IP addresses at a high rate; (2)  {\sc \bf DNS amplification} 
 using {\tt OpenDNS} server with {\tt BIND} (version 9.8) and emulating an  attacker
sending DNS requests with spoofed source IPs; (3) We  use {\tt iperf}
to create some fixed bandwidth traffic to generate {\sc \bf elephant flows}, and 
(4) {\sc \bf UDP flood} attacks.  We randomly pick one  edge PoP as the 
target and vary the target across  runs. We ramp up the attack volume  until
it induces maximum reduction in throughput of benign flows to the
target.  On our testbed, we can ramp up the volume  up to 10~Gbps. 
 For larger attacks, we use simulations.

\end{packeditemize}

\subsection{\Name scalability}
\label{subsec:benefits}


\mypara{Resource management}
Table~\ref{tab:res_manager_timeandopt} compares  the run time and optimality of the  
ILP-based algorithm and  \Name (i.e., DSP and SSP) for  3  ISP topologies of various sizes.
(We have results for several other  topologies but do not show it for brevity.) The ILP 
approach  takes from several tens of minutes to hours, whereas \Name takes only a 
few milliseconds  enabling rapid response
to changing traffic patterns.  The  optimality gap is $\leq$ 0.04\%.


 \begin{table}[t]
  \begin{center}
  \begin{small}
  \begin{tabular}{l|p{0.8cm}|l|l|l}
         Topology &  \#Nodes	&  \multicolumn{2}{c|}{Run time (secs)}  & Optimality \\ 
		& 		&  Baseline & \Name 	 &    Gap \\ \hline
 	Heanet  & 6 & 205 & 0.002  & 0.0003  \\  
	OTEGlobe & 92 &  2234  & 0.007   &  0.0004 \\
 	Cogent & 196  &  $>$ 1 hr & 0.01  & 0.0005 
     \end{tabular}
  \end{small}
  \end{center}
\vspace{-0.3cm}
 \tightcaption{Run time  and optimality gap of \Name vs.\ ILP formulation across different topologies.}
 \label{tab:res_manager_timeandopt}
\vspace{-0.1cm}
 \end{table}

\comment
{
 \begin{table}[t]
  \begin{center}
  \begin{footnotesize}
  \begin{tabular}{p{2.4cm}|p{1cm}|p{1cm}|p{1.1cm}|p{0.7cm}}
         Topo.(\# of switches)     & Heanet (6)  & ATT (24) 		& OTEGlobe (92) 		&Cogent (196)\\ \hline
          ILP (s)              	 &  205            & 452	& 2,234		& $>$1 hour	             \\ \hline
          \Name (s)       		&  0.002           & 0.003	& 0.007		& 0.010	             \\
     \end{tabular}
  \end{footnotesize}
  \end{center}
 \tightcaption{Run time of different resource management approaches across different topologies.}
 \label{tab:res_manager_time}
 \end{table}

 \begin{table}[t]
  \begin{center}
  \begin{footnotesize}
  \begin{tabular}{p{2.4cm}|p{1cm}|p{1cm}|p{1.1cm}|p{0.7cm}}
         Topo.(\# of switches)     & Heanet (6)  & ATT (24) 		& OTEGlobe (92) 		&Cogent (196)\\ \hline
          Optimality loss of \Name w.r.t. ILP (\%)  &  0.03\%            & 0.04\%	& 0.05\%		& 0.04\%	             \\
     \end{tabular}
  \end{footnotesize}
  \end{center}
 \tightcaption{Optimality loss of \Name over ILP.}
 \label{tab:res_manager_opt}
 \end{table}
}

\begin{figure}[t]
  \centering \includegraphics[width=200pt]{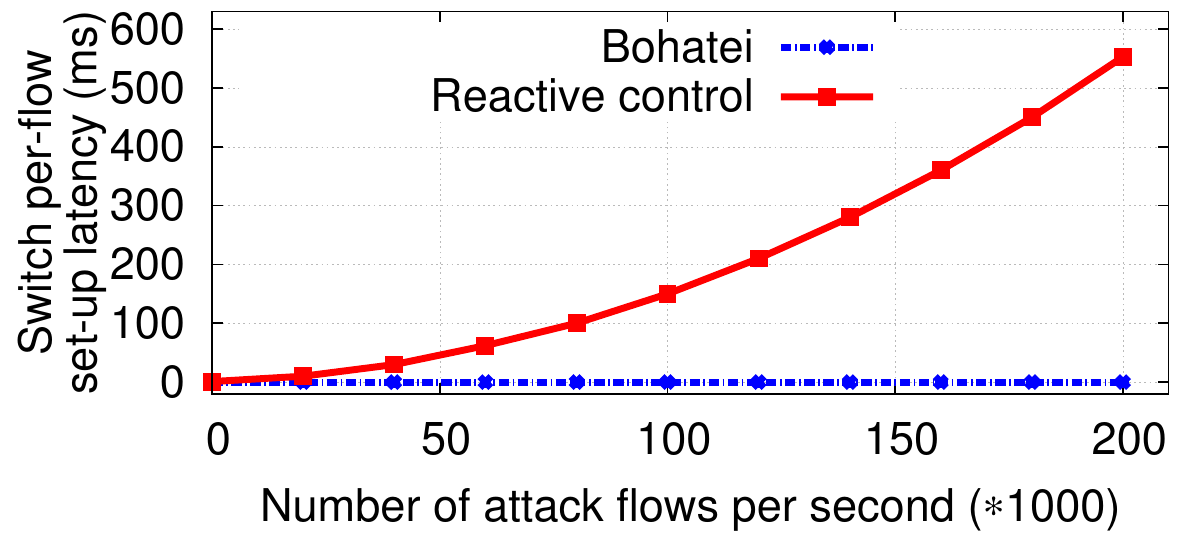}
  \tightcaption{\Name control plane scalability.}
	\vspace{-0.1cm}
\label{fig:control_scale}
 \end{figure}

\mypara{Control plane responsiveness} Figure~\ref{fig:control_scale} shows the
per-flow setup latency  comparing \Name to the  SDN per-flow and
reactive paradigm as the number of attack flows in a DNS amplification 
 attack increases. (The results are
consistent for other types of attacks and are not shown for brevity.)
 In both cases, we
have a dedicated machine for the controller with 8 2.8GHz cores and 64 GB RAM.
 To put the number of flows in context, 200K flows roughly corresponds to a  1~Gbps
attack.  Note that a typical upper bound for switch flow set-up time  is on
the order of a few milliseconds~\cite{amin_sdn_control}.
  We see that 
\Name  incurs zero rule setup latency, while 
the reactive approach deteriorates rapidly as the  attack volume increases.




\begin{figure}[t]
  \centering \includegraphics[width=200pt]{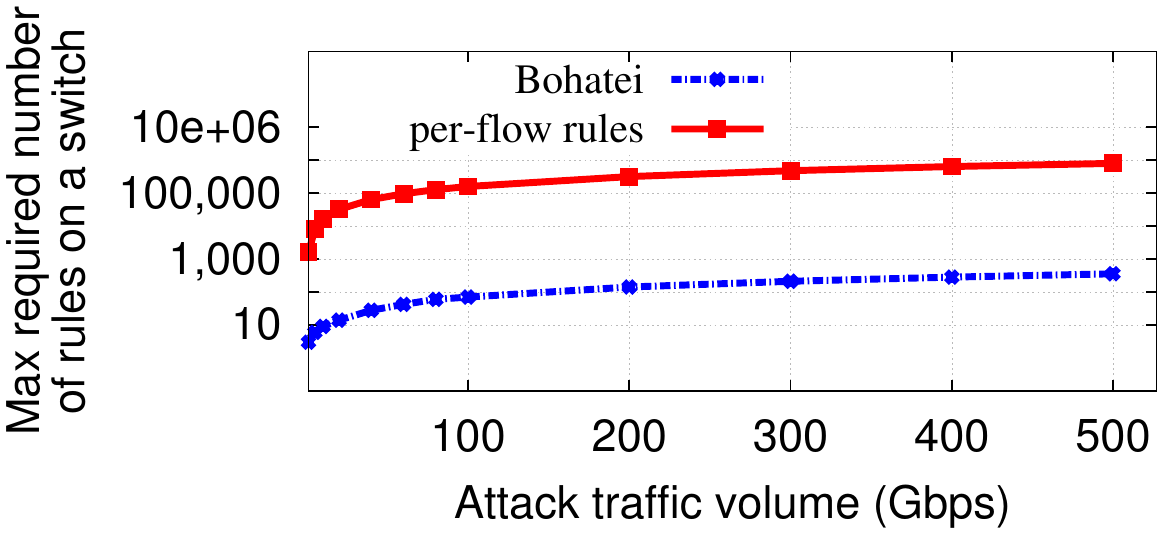}
 \vspace{-0.2cm}
  \tightcaption{Number of switch forwarding rules in \Name vs. 
  today's flow-based forwarding.\shepherd{A-7}}
 \vspace{-0.2cm}
\label{fig:sw_rules}
 \end{figure}

 \mypara{Number of forwarding rules} 
Figure~\ref{fig:sw_rules} shows the maximum number of rules required on 
a switch across different topologies for the SYN flood attack. Using today's 
flow-based forwarding, each new flow  will require a rule. Using tag-based
forwarding, the number of rules depends on the number of VM instances,
which reduces the switch rule space by four orders of magnitude. For other 
attack types, we observed consistent results (not shown).
To put this in context, the typical capacity of an SDN switch is
 3K-4K  rules (shared across various network management tasks).
 This means  that  per-flow rules will not suffice for attacks beyond
10Gbps. In contrast,  \Name can handle hundreds of Gbps of attack 
traffic; e.g., a 1~Tbps attack will require $<$ 1K  rules 
on a switch.


\mypara{Benefit of scale-out load balancing} We measured  the resources that
would be consumed by a  dedicated load balancing solution.  Across different
types of attacks with a fixed rate of 10Gbps, we observed that a dedicated load
balancer design requires between 220--300 VMs for load balancing alone. By
delegating the load balancing task to the VMs, our design obviates the need for
these extra load balancers (not shown). 

\comment{
\mypara{Benefit of rule push to edge} \vyas{TBD}
 Figure~\ref{fig:learning:eval} shows 
the effect of the optimization to  bypass analysis modules 
as soon as possible (\Section\ref{sec:otheropt}). Here, we consider  
a 180Mbps UDP flood attack that starts at $t=18s$ and is sustained 
until  $t=100s$). The figure shows that there is a 
significant gap in goodput between the cases of learning. Taking 
a closer look into this, we realized this is because the heavy CPU 
requirements of the analysis module. This is typical of many 
attack types: analysis tends to consume more computation power 
as compared  to response (e.g., just logging or dropping traffic). 
Using the learning optimization, such heavy operations are skipped 
once they are not needed any longer.

\begin{figure}[th]
  \vspace{-0.5cm}
  \centering \includegraphics[width=200pt]{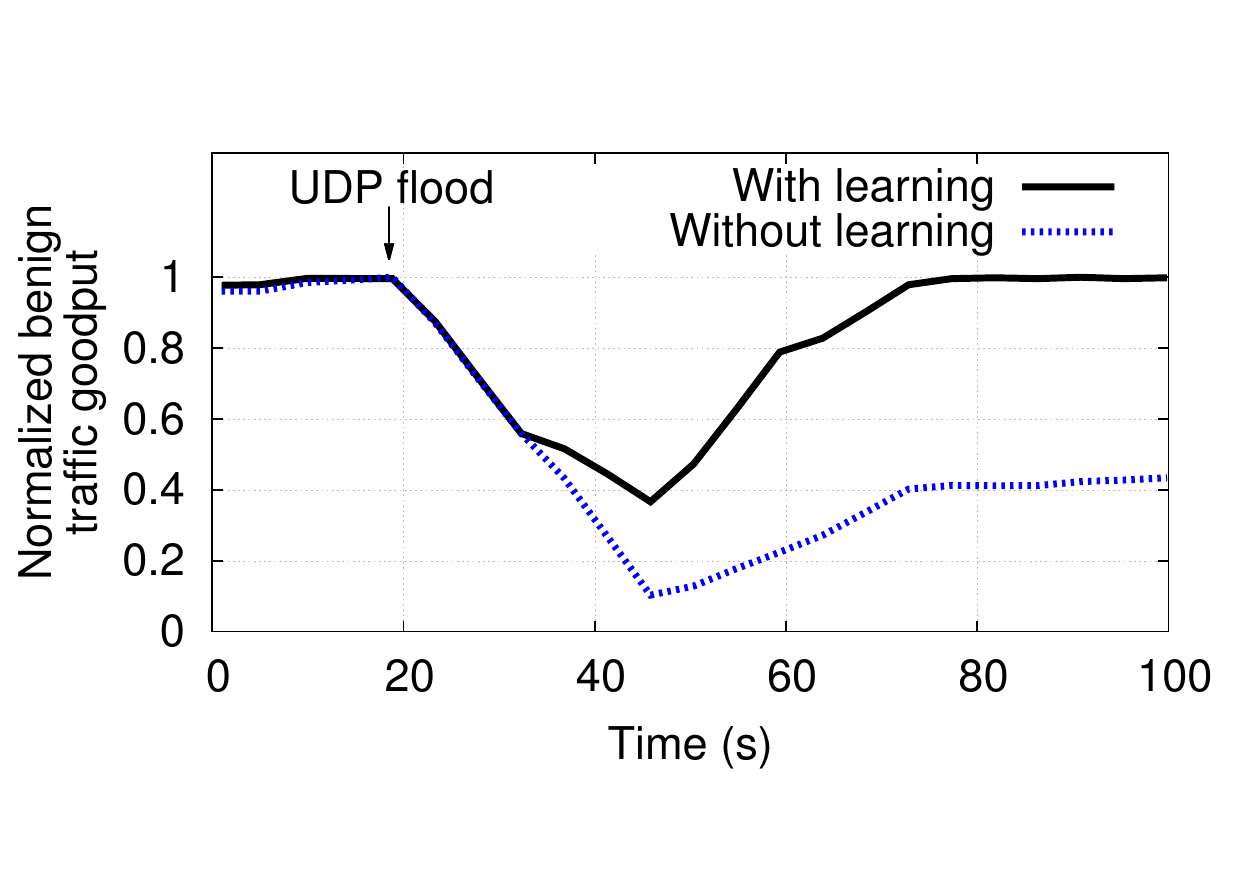}
  \vspace{-0.5cm}
  \tightcaption{Effect of the  optimization to preemptively push filtering rules to the edge.}
\label{fig:learning:eval}
 \end{figure}
}

\subsection{\Name end-to-end effectiveness}
\label{subsec:validation}

\begin{figure}[th]
  \vspace{-0.2cm}
  \centering \includegraphics[width=225pt]{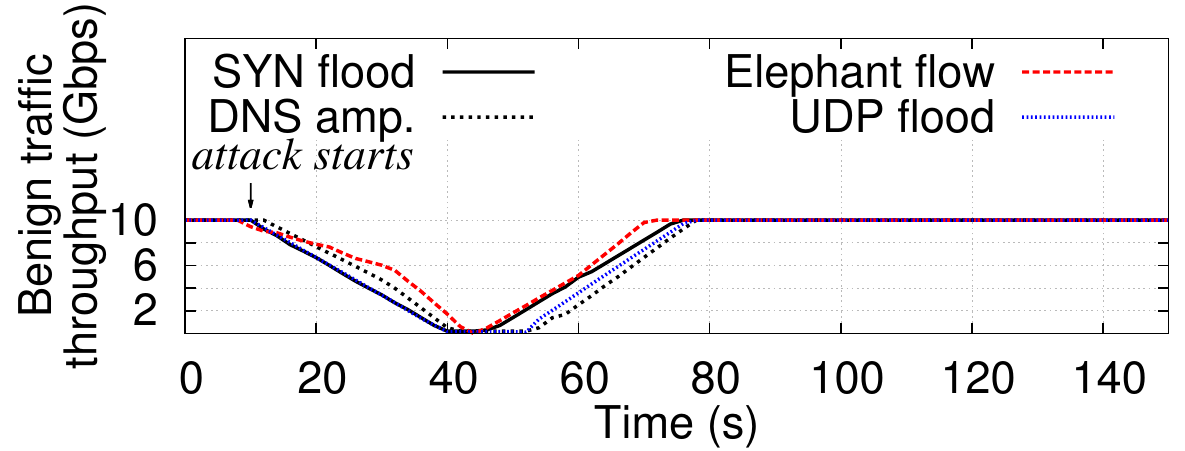}
  \vspace{-0.5cm}
  \tightcaption{\Name enables rapid response and restores 
    throughput  of legitimate traffic.}
\label{fig:e2e_different_attacks}
 \end{figure}

We  evaluated the effectiveness of \Name under four different types of \ddos
attacks.  We launch the attack traffic of the corresponding type at 10th 
second; the attack is sustained for the  duration of the experiment. In each 
scenario, we choose the attack volume such that it is capable of bringing 
the throughput of the benign traffic to zero. Figure~\ref{fig:e2e_different_attacks} 
shows the impact of attack traffic on the throughput of benign traffic. The Y axis 
for each scenario shows the network-wide throughput for TCP traffic (a total of  
10Gbps  if there is no attack).  The results shown in this figure are based on 
Cogent, the largest topology with 196 switches; the results for other topologies 
were consistent and are not shown. While we do see some small differences   
across attacks, the overall reaction time is short.

The key takeaway is that  \Name can help networks respond rapidly  (within one
minute) to diverse attacks  and restore the performance of legitimate  flows.
We repeated the experiments with UDP as the benign traffic. In this case,
the recovery time was even shorter, as the throughput does not suffer from 
the congestion control mechanism of TCP.




 \begin{table}[t]
  \begin{center}
  \begin{footnotesize}
  \begin{tabular}{l|l|l}
         Attack type 	& \multicolumn{2}{c}{\# VMs needed} \\ 
		&  Monolithic& Fine-grained scaling \\ \hline 
	DNS Amplification	& 5,422 & 1,005 \\
	SYN Flood	& 3,167 & 856 \\
	Elephant flows & 1,948 & 910 \\
	UDP flood & 3,642 &  1,253
     \end{tabular}
  \end{footnotesize}
  \end{center}
 \vspace{-0.2cm}
 \tightcaption{Total hardware provisioning cost needed to handle a 100~Gbps attack 
 for different attacks.}
 \vspace{-0.2cm}
 \label{tab:finescale}
 \end{table}

 \mypara{Hardware cost} We measure the total number of VMs 
 needed to handle  a given attack volume and compare two cases:
 (1)  monolithic  VMs  embedding the entire defense logic for an attack, 
 and (2) using \Name's fine-grained modular scaling.  Table~\ref{tab:finescale}
shows the number of VMs required to handle different types of 100~Gbps attacks.
  Fine-grained scaling gives a 2.1--5.4$\times$ reduction in hardware cost 
  vs.\ monolithic VMs.  Assuming a commodity server costs  \$3,000 and  can run 
  40VMs in \Name (as  we did), we see that it takes a total hardware cost of 
  less than about \$32,000  to handle a 100~Gbps attack across 
  Table~\ref{tab:finescale}. This is in contrast to the total server cost of
about \$160,000 for the same scenario if we use monolithic VMs. Moreover, 
since \Name is horizontally scalable by construction, dealing with
larger attacks simply entails a linearly scale up of the number of VMs. 


\begin{figure}[t]
  \centering \includegraphics[width=200pt]{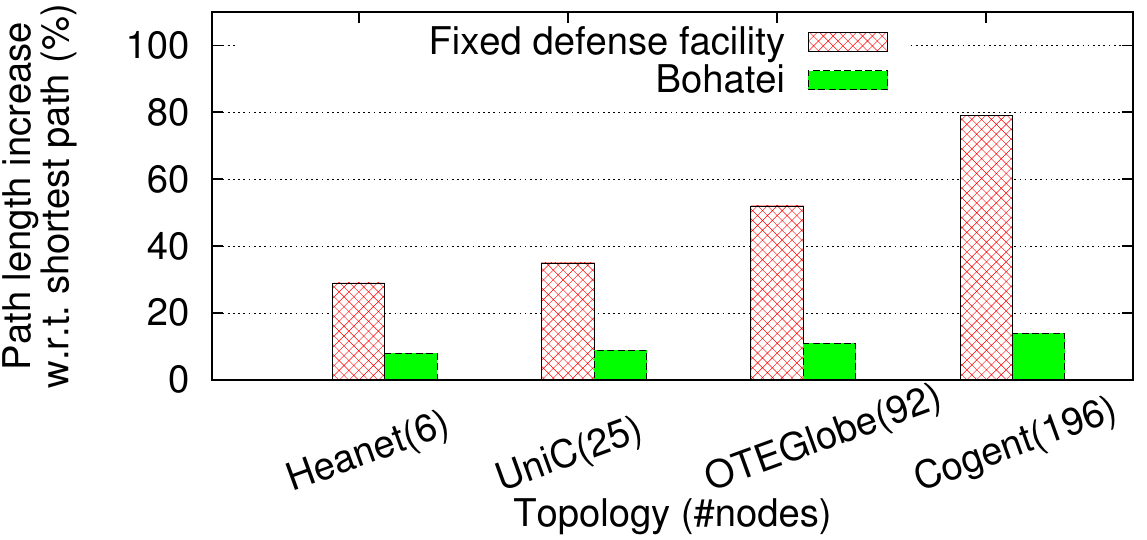}
  \tightcaption{Routing efficiency in \Name.}
\vspace{-0.1cm}
\label{fig:routing_ineff}
 \end{figure}

\mypara{Routing efficiency} To  quantify how  \Name addresses the routing
inefficiency of existing solutions (\Section\ref{subsec:opportunities}), we 
ran the following experiment.  For each  topology, we  measured the end-to-end 
latency in two equivalently provisioned scenarios: (1) the location of the \ddos 
defense appliance is the node with the highest betweenness 
value\footnote{Betweenness is a measure of a node's centrality, which is the 
fraction of the network's all-pairs shortest paths that pass through that node.}, 
and (2)  \Name.  As a baseline, we consider shortest path routing without attacks. 
The  main conclusion in Figure~\ref{fig:routing_ineff} is that \Name reduces 
traffic latency by 20\% to 65\% across different scenarios.


\subsection{Dynamic \ddos attacks}
\label{subsec:advesary_eval}

We consider the following dynamic \ddos attack strategies:
(1) \emph{RandIngress}: In each epoch, pick a random subset of attack 
ingresses and  distribute the attack budget evenly across attack types; (2)
\emph{RandAttack}: In each epoch, pick a random subset of attack types and
distribute the budget evenly across all ingresses; (3)  \emph{RandHybrid}: In
each epoch, pick a random subset of ingresses and  attack types independently
and  distribute  the attack budget evenly across selected pairs; (4)
\emph{Steady}:  The adversary picks a random attack type and a subset of ingresses
and sustains it during all epochs; and (5)  \emph{FlipPrevEpoch}: This is conceptually
equivalent to conducting two \emph{Steady} attacks $A1$ and $A2$ with each 
being active during odd and even epochs, respectively.

 \begin{figure}[t]
\begin{center}
\subfloat[Regret w.r.t. defense resource consumption.]
{
  \includegraphics[width=220pt]{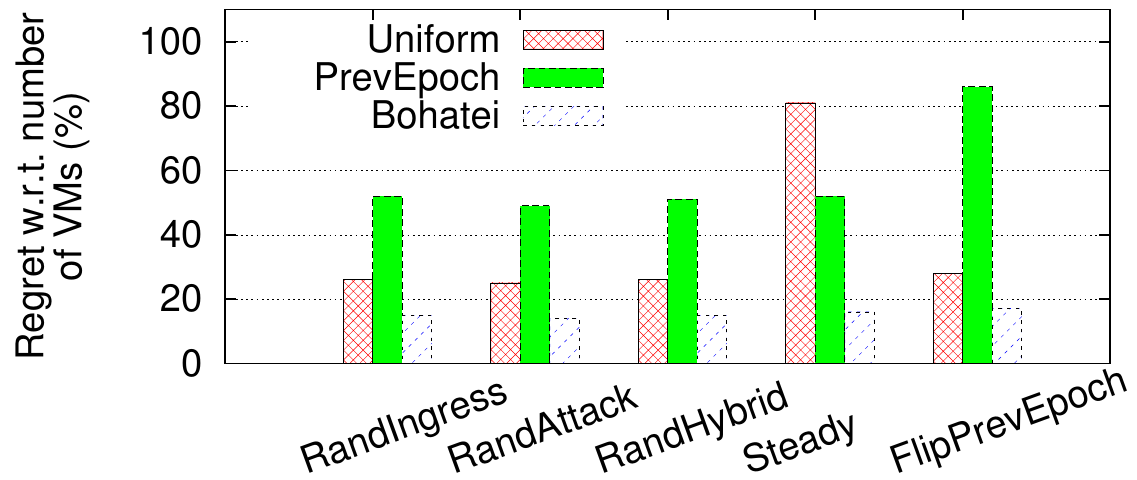}
\label{fig:cost_vms}
} \\ \vspace{-0.2cm}
\subfloat[Regret w.r.t. successful attacks.]
 { \includegraphics[width=220pt]{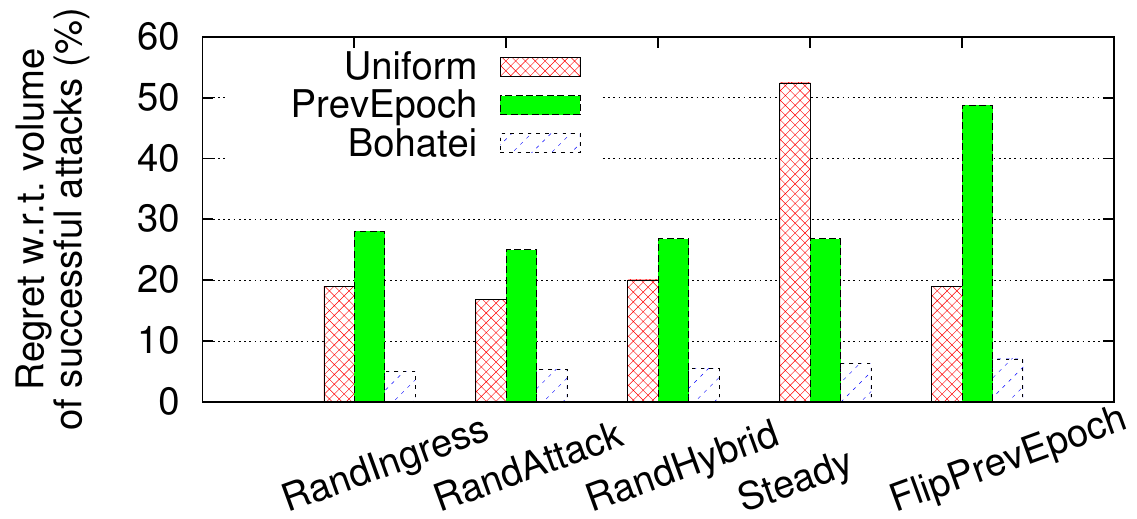}
\label{fig:cost_attack}
} \\ \vspace{-0.2cm}
\end{center}
\tightcaption{Effect of different adaptation strategies (bars) vs. different attacker 
strategies (X axis).}
\label{fig:online_adaptation}
 \vspace{-0.4cm}
 \end{figure}

 Given the typical \ddos attack duration ($\approx$ 6 hours~\cite{incapsula}), we
consider an attack lasting for 5000 5-second epochs (i.e., $\approx$7 hours).
\Name is initialized with a zero starting point of attack estimates.  The
metric of interest we report is the {\em normalized regret} with respect to the
best static  decision  in hindsight; i.e., if we had to pick a single static
strategy for the entire duration.  Figure~\ref{fig:cost_vms} and
Figure~\ref{fig:cost_attack} show the regret w.r.t.\ the two goals G1 
(the number of VMs) and G2 (volume of successful attack) for a 24-node 
topology.  The results are similar using other topologies and are not shown here. 
Overall, \Name's online adaptation  achieves low regret across the  adversarial 
strategies compared to two strawman solutions: (1)  uniform  estimates, and (2) 
estimates given the previous measurements.

\section{Related Work}
\label{sec:relwork}

\ddos has a long history; we refer readers to surveys for a taxonomy of \ddos 
attacks and defenses (e.g.,~\cite{reihersurvey}).  We 
have already discussed relevant SDN/NFV work in the previous sections.  
Here, we briefly review other related topics.




\mypara{Attack detection} There are several algorithms for detecting and 
filtering DDoS attacks.  These include time series detection techniques 
(e.g.,~\cite{wavelet}), use  of backscatter analysis 
(e.g.,~\cite{Moore:2006:IID:1132026.1132027}), exploiting attack-specific 
features (e.g.,~\cite{Jin:2003:HFE:948109.948116}), and network-wide
analysis (e.g.,~\cite{Lakhina05:Sigcomm}).  These are orthogonal to
the focus of this paper.

\myparatight{\ddos-resilient Internet architectures} These include the use of 
capabilities~\cite{tva}, better inter-domain routing
(e.g.,~\cite{scion}), inter-AS collaboration
(e.g.,~\cite{pushback}),  packet marking and  unforgeable
identifiers (e.g.,~\cite{aip}), and traceback (e.g.,~\cite{savage_traceback}). 
However, they do not provide an immediate deployment path 
or resolution for current networks.  In contrast, \Name focuses on a more 
practical, single-ISP context, and is aligned
with economic incentives for ISPs and their  customers.

\myparatight{Overlay-based solutions}  There are 
overlay-based solutions (e.g.,~\cite{overdose,mayday})
 that act as a  ``buffer zone'' between attack sources and targets. 
The design contributions in  \Name can be applied to these as well.

%

\myparatight{SDN/NFV-based security} There are few  efforts in this space 
such as FRESCO~\cite{fresco} and AvantGuard~\cite{avantguard}.
As we saw earlier, these SDN  solutions
will introduce new DDoS avenues because of the per-flow and reactive
model~\cite{avantguard}. Solving this control bottleneck requires hardware
modifications to SDN switches to add ``stateful'' components, which is unlikely
to be supported by switch vendors soon~\cite{avantguard}.  In contrast, \Name 
chooses a proactive approach of setting up tag-based forwarding rules that is immune to
these pitfalls. 




\section{Conclusions}
\label{sec:conc}
 
\Name  brings the flexibility and elasticity benefits of recent networking trends, 
such as SDN and NFV, to \ddos defense. We 
addressed practical challenges  in the design of \Name's
resource management algorithms and  control/data plane mechanisms to ensure
that these do not become bottlenecks for \ddos defense.  We implemented a
full-featured \Name prototype built on industry-standard SDN control platforms
and commodity network appliances. Our evaluations on a real testbed show that
\Name (1) is scalable and responds rapidly to attacks, (2) outperforms naive
SDN implementations that do not address the control/data plane bottlenecks, and (3)
enables resilient defenses against  dynamic adversaries.
 Looking forward, we believe that  these design principles can also be applied 
to other aspects of network security.

\section*{Acknowledgments}
This work was supported in part by grant number N00014-13-1-0048 from
the Office of Naval Research, and NSF awards 1409758, 1111699,
1440056, and 1440065. Seyed K. Fayaz was supported in part by the CMU
Bertucci Fellowship. We thank Limin Jia, Min Suk Kang, the anonymous 
reviewers, and our shepherd Patrick Traynor for their helpful suggestions.

{
\scriptsize
\bibliographystyle{abbrv}
\bibliography{bibs/seyed,bibs/smartre,bibs/mboxsdn,bibs/aplomb,bibs/mbox,bibs/paper,bibs/bailey,bibs/csamp.bib}

\begin{thebibliography}{10}

\bibitem{bohatei_code}
{Bohatei}.
\newblock \url{https://github.com/ddos-defense/bohatei}.

\bibitem{ec2}
{Amazon EC2}.
\newblock \url{http://aws.amazon.com/ec2/}.

\bibitem{arbor}
{Arbor Networks}, worldwide infrastructure security report, volume {IX}, 2014.
\newblock \url{http://bit.ly/1R0NDRi}.

\bibitem{att_intel}
{AT\&T and Intel: Transforming the Network with NFV and SDN}.
\newblock \url{https://www.youtube.com/watch?v=F55pHxTeJLc#t=76}.

\bibitem{attprotect}
{AT\&T Denial of Service Protection}.
\newblock \url{http://soc.att.com/1IIlUec}.

\bibitem{att_vision}
{AT\&T Domain 2.0 Vision White Paper}.
\newblock \url{http://soc.att.com/1kAw1Kp}.

\bibitem{click}
{Click Modular Router}.
\newblock \url{http://www.read.cs.ucla.edu/click/click}.

\bibitem{cloudflare}
{CloudFlare}.
\newblock \url{https://www.cloudflare.com/ddos}.

\bibitem{synproxy_redhat}
{DDoS protection using Netfilter/iptables}.
\newblock \url{http://bit.ly/1IImM2F}.

\bibitem{dell_server}
{Dell PowerEdge Rack Servers}.
\newblock \url{http://dell.to/1dtP5Jk}.

\bibitem{gsa_schedules}
{GSA Advantage}.
\newblock \url{http://1.usa.gov/1ggEgFN}.

\bibitem{incapsula}
{Incapsula Survey : What DDoS Attacks Really Cost Businesses, 2014}.
\newblock \url{http://bit.ly/1CFZyIr}.

\bibitem{iptables}
{iptables}.
\newblock \url{http://www.netfilter.org/projects/iptables/}.

\bibitem{hockey_stick_era}
{NTP} attacks: Welcome to the hockey stick era.
\newblock \url{http://bit.ly/1ROlwQe}.

\bibitem{ons_14_keynote}
{ONS 2014 Keynote: John Donovan, Senior EVP, AT\&T Technology \& Network
  Operations}.
\newblock \url{http://bit.ly/1RQFMko}.

\bibitem{openvswitch}
{Open vSwitch}.
\newblock \url{http://openvswitch.org/}.

\bibitem{odl}
{OpenDaylight} project.
\newblock \url{http://www.opendaylight.org/}.

\bibitem{dpdk}
{Packet processing on Intel architecture}.
\newblock \url{http://intel.ly/1efIEu6}.

\bibitem{prolexic}
Prolexic.
\newblock \url{http://www.prolexic.com/}.

\bibitem{radware}
{Radware}.
\newblock \url{http://www.radware.com/Solutions/Security/}.

\bibitem{sdn_sequel}
{Time for an SDN Sequel?}
\newblock \url{http://bit.ly/1BSpdma}.

\bibitem{topology_zoo}
{Topology Zoo}.
\newblock \url{www.topology-zoo.org}.

\bibitem{verizon_sdn}
{Verizon-Carrier Adoption of Software-defined Networking}.
\newblock \url{https://www.youtube.com/watch?v=WVczl03edi4}.

\bibitem{zscalar}
{ZScaler Cloud Security}.
\newblock \url{http://www.zscaler.com}.

\bibitem{mayday}
D.~G. Andersen.
\newblock Mayday: Distributed filtering for internet services.
\newblock In {\em Proc.\ {USITS}}, 2003.

\bibitem{aip}
D.~G. Andersen, H.~Balakrishnan, N.~Feamster, T.~Koponen, D.~Moon, and
  S.~Shenker.
\newblock Accountable internet protocol {(AIP)}.
\newblock In {\em Proc.\ SIGCOMM}, 2008.

\bibitem{wavelet}
P.~Barford, J.~Kline, D.~Plonka, and A.~Ron.
\newblock A signal analysis of network traffic anomalies.
\newblock In {\em Proc.\ {ACM SIGCOMM Workshop on Internet Measurement}}, 2002.

\bibitem{synproxy_paper}
R.~C\'{a}ceres, F.~Douglis, A.~Feldmann, G.~Glass, and M.~Rabinovich.
\newblock Web proxy caching: The devil is in the details.
\newblock {\em SIGMETRICS Perform. Eval. Rev.}, 26(3):11--15, Dec. 1998.

\bibitem{fabric}
M.~Casado, T.~Koponen, S.~Shenker, and A.~Tootoonchian.
\newblock Fabric: A retrospective on evolving sdn.
\newblock In {\em Proc.\ {HotSDN}}, 2012.

\bibitem{gorilla}
J.~Czyz, M.~Kallitsis, M.~Gharaibeh, C.~Papadopoulos, M.~Bailey, and M.~Karir.
\newblock Taming the 800 pound gorilla: The rise and decline of ntp ddos
  attacks.
\newblock In {\em Proc. IMC}, 2014.

\bibitem{flowtags_nsdilong}
S.~K. Fayazbakhsh, L.~Chiang, V.~Sekar, M.~Yu, and J.~C. Mogul.
\newblock Enforcing network-wide policies in the presence of dynamic middlebox
  actions using {FlowTags}.
\newblock In {\em Proc. NSDI}, 2014.

\bibitem{4d}
A.~Greenberg, G.~Hjalmtysson, D.~A. Maltz, A.~Myers, J.~Rexford, G.~Xie,
  H.~Yan, J.~Zhan, and H.~Zhang.
\newblock A clean slate {4D} approach to network control and management.
\newblock {\em ACM CCR}, 2005.

\bibitem{snips_iciss}
V.~Heorhiadi, S.~K. Fayaz, M.~Reiter, and V.~Sekar.
\newblock Frenetic: A network programming language.
\newblock {\em Information Systems Security}, 2014.

\bibitem{google_sdn}
{Jain et~al}.
\newblock B4: Experience with a globally-deployed software defined wan.
\newblock In {\em Proc. SIGCOMM}, 2013.

\bibitem{Jin:2003:HFE:948109.948116}
C.~Jin, H.~Wang, and K.~G. Shin.
\newblock Hop-count filtering: An effective defense against spoofed ddos
  traffic.
\newblock In {\em Proc. CCS}, 2003.

\bibitem{kalai}
A.~Kalai and S.~Vempala.
\newblock Efficient algorithms for online decision problems.
\newblock {\em J. Comput. Syst. Sci.}, 2005.

\bibitem{crossfire}
M.~S. Kang, S.~B. Lee, and V.~Gligor.
\newblock The crossfire attack.
\newblock In {\em Proc.\ {IEEE Security and Privacy}}, 2013.

\bibitem{Lakhina05:Sigcomm}
A.~Lakhina, M.~Crovella, and C.~Diot.
\newblock {Mining Anomalies Using Traffic Feature Distributions}.
\newblock In {\em Proc. SIGCOMM}, 2005.

\bibitem{pushback}
R.~Mahajan et~al.
\newblock Controlling high bandwidth aggregates in the network.
\newblock {\em CCR}, 2001.

\bibitem{openflow}
N.~McKeown et~al.
\newblock {OpenFlow}: enabling innovation in campus networks.
\newblock {\em CCR}, March 2008.

\bibitem{reihersurvey}
J.~Mirkovic and P.~Reiher.
\newblock A taxonomy of ddos attack and ddos defense mechanisms.
\newblock In {\em CCR}, 2004.

\bibitem{Moore:2006:IID:1132026.1132027}
D.~Moore, C.~Shannon, D.~J. Brown, G.~M. Voelker, and S.~Savage.
\newblock Inferring internet denial-of-service activity.
\newblock {\em ACM Trans. Comput. Syst.}, 2006.

\bibitem{etsinfv}
Network functions virtualisation -- introductory white paper.
\newblock \url{http://portal.etsi.org/NFV/NFV_White_Paper.pdf}.

\bibitem{arbor-number}
A.~Networks.
\newblock {ATLAS Summary Report: Global Denial of Service}.
\newblock \url{http://atlas.arbor.net/summary/dos}.

\bibitem{ananta}
P.~Patel et~al.
\newblock {Ananta: cloud scale load balancing}.
\newblock In {\em Proc.\ ACM SIGCOMM}, 2013.

\bibitem{bro}
V.~Paxson.
\newblock Bro: A system for detecting network intruders in real-time.
\newblock In {\em Computer Networks}, 1999.

\bibitem{arrakis}
S.~Peter, J.~Li, I.~Zhang, D.~R.~K. Ports, D.~Woos, A.~Krishnamurthy,
  T.~Anderson, and T.~Roscoe.
\newblock Arrakis: The operating system is the control plane.
\newblock In {\em Proc.\ {OSDI}}, 2014.

\bibitem{snort}
M.~Roesch.
\newblock {Snort - Lightweight Intrusion Detection for Networks}.
\newblock In {\em LISA}, 1999.

\bibitem{ntp_hell}
C.~Rossow.
\newblock Amplification hell: Revisiting network protocols for ddos abuse.
\newblock In {\em Proc. USENIX Security}, 2014.

\bibitem{r:05}
M.~Roughan.
\newblock {Simplifying the Synthesis of Internet Traffic Matrices}.
\newblock {\em ACM SIGCOMM CCR}, 2005.

\bibitem{savage_traceback}
S.~Savage, D.~Wetherall, A.~Karlin, and T.~Anderson.
\newblock Practical network support for ip traceback.
\newblock In {\em Proc. SIGCOMM}, 2000.

\bibitem{overdose}
E.~Shi, I.~Stoica, D.~Andersen, and A.~Perrig.
\newblock {OverDoSe}: A generic {DDoS} protection service using an overlay
  network.
\newblock Technical Report CMU-CS-06-114, School of Computer Science, Carnegie
  Mellon University, 2006.

\bibitem{fresco}
S.~Shin, P.~Porras, V.~Yegneswaran, M.~Fong, G.~Gu, and M.~Tyson.
\newblock {FRESCO}: Modular composable security services for software-defined
  networks.
\newblock In {\em Proc.\ {NDSS}}, 2013.

\bibitem{avantguard}
S.~Shin, V.~Yegneswaran, P.~Porras, and G.~Gu.
\newblock {AVANT-GUARD}: Scalable and vigilant switch flow management in
  software-defined networks.
\newblock In {\em Proc.\ {CCS}}, 2013.

\bibitem{coremelt}
A.~Studer and A.~Perrig.
\newblock The coremelt attack.
\newblock In {\em Proc.\ {ESORICS}}, 2009.

\bibitem{vmware}
{T. Koponen et~al}.
\newblock Network virtualization in multi-tenant datacenters.
\newblock In {\em Proc. NSDI}, 2014.

\bibitem{irscp-usenix07}
P.~Verkaik, D.~Pei, T.~Schollf, A.~Shaikh, A.~C. Snoeren, and J.~E. van~der
  Merwe.
\newblock {Wresting Control from BGP: Scalable Fine-grained Route Control}.
\newblock In {\em Proc.\ USENIX ATC}, 2007.

\bibitem{tva}
X.~Yang, D.~Wetherall, and T.~Anderson.
\newblock A dos-limiting network architecture.
\newblock In {\em Proc. SIGCOMM}, 2005.

\bibitem{amin_sdn_control}
S.~Yeganeh, A.~Tootoonchian, and Y.~Ganjali.
\newblock On scalability of software-defined networking.
\newblock {\em Communications Magazine, IEEE}, 2013.

\bibitem{scion}
X.~Zhang, H.-C. Hsiao, G.~Hasker, H.~Chan, A.~Perrig, and D.~G. Andersen.
\newblock Scion: Scalability, control, and isolation on next-generation
  networks.
\newblock In {\em Proc.\ {IEEE Security and Privacy}}, 2011.

\end{thebibliography}
}


\appendix 


\section{ILP Formulation}
\label{subsec:ilp}
\vspace{-2mm}

\begin{figure*}[t!]
\begin{small}
  \fbox{
  \begin{minipage}{0.97\linewidth}

\begin{codebox}
\li Minimize
$\alpha
\times 
\sum\limits_{\edgepopindex}^{}
\sum\limits_{\attackindex}^{}
\sum\limits_{\datacenterindex}^{}
\fraction_{\edgepopindex,\attackindex,\datacenterindex}
\times
\Traffic_{\edgepopindex,\attackindex}
\times
\cost_{\edgepopindex,\datacenterindex}
+
\sum\limits_{\datacenterindex}^{}
\datacentercost_{\datacenterindex}
$

\\ s.t.

\li $ \forall  \edgepopindex, \attackindex:
\sum\limits_{\datacenterindex}^{}
\fraction_{\edgepopindex,\attackindex,\datacenterindex} = 1$
\Comment all suspicious traffic should be served

\li $ \forall \attackindex, \datacenterindex:
\atraffic_{\attackindex, \datacenterindex}
=
\sum\limits_{\edgepopindex}^{}
\fraction_{\edgepopindex,\attackindex,\datacenterindex}
\times
\Traffic_{\edgepopindex,\attackindex}$
\Comment traffic of each type to each datacenter

\li $ \forall \datacenterindex:
\sum\limits_{\attackindex}\atraffic_{\attackindex, \datacenterindex}
\leq
\capacity^{link}_{\datacenterindex}$
\Comment datacenter link capacity

\li $ \forall \datacenterindex, \attackindex, \vertexindexi:
\sum\limits_{\aserver \in \servers_\datacenterindex}
\vms_{\attackindex,\vertexindexi}^{\datacenterindex,\aserver} 
\geq
\atraffic_{\attackindex, \datacenterindex} 
\times
\frac{\sum\limits_{\vertexindexj: (\vertexindexj,\vertexindexi) = \edge_{\attackindex,\vertexindexj \rightarrow \vertexindexi}^{\logical}}{}{\weight_{\attackindex,\vertexindexj \rightarrow \vertexindexi}}}{\power_{\attackindex,\vertexindexi}}$
\Comment provisioning sufficient VMs ($\servers_\datacenterindex$ 
is the set of $\datacenterindex$'s servers.)

\li $\forall \datacenterindex, \aserver \in \servers_\datacenterindex :
\sum\limits_{\attackindex}
\sum\limits_{\vertexindexi}
\vms_{\attackindex,\vertexindexi}^{\datacenterindex,\aserver} 
\leq
\capacity^{compute}_{\datacenterindex,\aserver}$
\Comment server compute capacity

\li $ \forall \datacenterindex:
\datacentercost_\datacenterindex = 
\intrarack_\datacenterindex 
\times
\unitIntraRackCost
+ 
\interrack_\datacenterindex
\times  
\unitInterRackCost
$
\Comment total cost within each datacenter

\li $ \forall \datacenterindex:
\intrarack_{\datacenterindex} =
\sum\limits_{\attackindex}
\sum\limits_{(\vertexindexi,\vertexindexj) = 
\edge_{\attackindex,\vertexindexi \rightarrow \vertexindexj}^{\logical}} 
\sum \limits_{(\aserver, \aserver') \in sameRack}
\sum \limits_{\vmindex=1}^{\maxVM}
\sum \limits_{\vmindex'=1}^{\maxVM}
\sum \limits_{\trafficunit=1}^{\maxVol}
\vmassigned_{\datacenterindex, \attackindex,\vertexindexi,\vmindex,\aserver, \vertexindexj, \vmindex',\aserver',\trafficunit}
$
\Comment intra-rack cost

\li $ \forall \datacenterindex:
\interrack_{\datacenterindex} =
\sum\limits_{\attackindex}
\sum\limits_{(\vertexindexi,\vertexindexj) = \edge_{\attackindex,\vertexindexi \rightarrow \vertexindexj}^{\logical}} 
\sum \limits_{(\aserver, \aserver') \notin sameRack}
\sum \limits_{\vmindex=1}^{\maxVM}
\sum \limits_{\vmindex'=1}^{\maxVM}
\sum \limits_{\trafficunit=1}^{\maxVol}
\vmassigned_{\datacenterindex, \attackindex,\vertexindexi,\vmindex,\aserver, \vertexindexj, \vmindex',\aserver',\trafficunit}
$
\Comment inter-rack cost

\li  $ \forall \datacenterindex, \attackindex, \vertexindexj, \vmindex':
\sum \limits_{\aserver}
\sum \limits_{\aserver'}
\sum\limits_{\vertexindexi: (\vertexindexi,\vertexindexj) = \edge_{\attackindex,\vertexindexi \rightarrow \vertexindexj}^{\logical}} 
\sum \limits_{\vmindex=1}^{\maxVM}
\sum \limits_{\trafficunit=1}^{\maxVol}
\vmassigned_{\datacenterindex, \attackindex,\vertexindexi,\vmindex,\aserver, \vertexindexj, \vmindex',\aserver',\trafficunit} \leq 
\power_{\attackindex,\vertexindexj}
$
\Comment enforcing VMs capacities

\li $ \forall \datacenterindex,
\aserver \in \servers_\datacenterindex, 
\attackindex, 
\vertexindexj:
\vms_{\attackindex,\vertexindexj}^{\datacenterindex,\aserver} 
\times
\power_{\attackindex,\vertexindexj} \geq 
\sum\limits_{\vmindex=1}^{\maxVM}
\sum \limits_{\vmindex'=1}^{\maxVM}
\sum\limits_{\vertexindexi: (\vertexindexi,\vertexindexj) = \edge_{\attackindex,\vertexindexi \rightarrow \vertexindexj}^{\logical}} \sum\limits_{\aserver'}
\sum \limits_{\trafficunit=1}^{\maxVol}
\vmassigned_{\datacenterindex, \attackindex,\vertexindexi,\vmindex,\aserver, \vertexindexj, \vmindex',\aserver',\trafficunit}
$
\Comment bound  traffic volumes

\li 
$ \forall \datacenterindex,
\aserver \in \servers_\datacenterindex, 
\attackindex, 
\vertexindexj:
\vms_{\attackindex,\vertexindexj}^{\datacenterindex,\aserver} 
\times
\power_{\attackindex,\vertexindexj}
\leq 
\sum\limits_{\vmindex=1}^{\maxVM}
\sum \limits_{\vmindex'=1}^{\maxVM}
\sum\limits_{\vertexindexi: (\vertexindexi,\vertexindexj) = \edge_{\attackindex,\vertexindexi \rightarrow \vertexindexj}^{\logical}} \sum\limits_{\aserver'}
\sum \limits_{\trafficunit=1}^{\maxVol}
\vmassigned_{\datacenterindex, \attackindex,\vertexindexi,\vmindex,\aserver, \vertexindexj, \vmindex',\aserver',\trafficunit} + 1
$
\Comment bound  traffic volumes

\li \Comment flow conservation for VM $\vmindex$ of type logical node $k$ that has both predecessor(s) and successor(s)
\\$ \forall \datacenterindex,
\attackindex, 
k, 
\vmindex:
\sum \limits_{\vmindex'=1}^{\maxVM}
\sum\limits_{g: (g,k) = \edge_{\attackindex,g \rightarrow k}^{\logical}} 
\sum\limits_{\aserver}
\sum\limits_{\aserver'}
\sum \limits_{\trafficunit=1}^{\maxVol}
\vmassigned_{\datacenterindex, \attackindex,g,\vmindex',\aserver', k, \vmindex,\aserver,\trafficunit}
=
\sum \limits_{\vmindex'=1}^{\maxVM}
\sum\limits_{h: (k,h) = \edge_{\attackindex,k \rightarrow h}^{\logical}} 
\sum\limits_{\aserver}
\sum\limits_{\aserver'}
\sum \limits_{\trafficunit=1}^{\maxVol}
\vmassigned_{\datacenterindex, \attackindex,k,\vmindex,\aserver, h, \vmindex',\aserver',\trafficunit}
$

\li $ \forall link \in ISP\;backbone:
\sum\limits_{link \in Path_{\edgepopindex \rightarrow \datacenterindex}}^{}
\sum\limits_{\attackindex}^{}
\fraction_{\edgepopindex,\attackindex,\datacenterindex}
\times
\Traffic_{\edgepopindex,\attackindex}
\leq
\beta \times MaxLinkCapacity
$
\Comment per-link traffic load control

\li $\fraction_{\edgepopindex,\attackindex,\datacenterindex} \in [0,1],
\vmassigned_{\datacenterindex, \attackindex,\vertexindexi,\vmindex,\aserver, \vertexindexj, \vmindex',\aserver',\trafficunit} \in \{0,1\}, \vms_{\attackindex,\vertexindexi}^\datacenterindex,
\vms_{\attackindex,\vertexindexi}^{\datacenterindex,\aserver} \in \{0,1,\dots\},
\atraffic_{\attackindex, \datacenterindex}, \interrack_{\datacenterindex},
\intrarack_{\datacenterindex}, \datacentercost_{\datacenterindex} \in \mathbb{R}
$
\Comment variables

\end{codebox}
  \end{minipage}
}
\end{small}
	\vspace{-0.3cm}
  \caption{ILP formulation for an optimal resource management.}
  \vspace*{3in}
	\label{fig:milp}
	\vspace{-8.2cm}
\end{figure*}
The ILP formulation for an optimal resource management 
(mentioned in \Section\ref{subsec:problem_statement_res}) is shown 
in Figure~\ref{fig:milp}. 

\mypara{Vairables} In addition to the parameters and variables that we 
have defined earlier in \Section\ref{sec:resource_layer}, we define the binary variable 
$\vmassigned_{\datacenterindex, \attackindex,\vertexindexi,\vmindex,\aserver, \vertexindexj, \vmindex',\aserver',\trafficunit}$ 
as follows: if  it is 1, VM $\vmindex$ of type $\vertex_{\attackindex,\vertexindexi}$ 
runs on  server $\aserver$ and sends 1 unit of traffic (e.g., 1 Gbps) to 
 VM $\vmindex'$ of type $\vertex_{\attackindex,\vertexindexj}$ that runs 
on  server $\aserver'$, where 
$\edge_{\attackindex,\vertexindexi \rightarrow \vertexindexj}^{\logical}
\in \edges^{\logical}_\attackindex$, and servers $\aserver$ and $\aserver'$ are
located in datacenter $\datacenterindex$; otherwise, 
$\vmassigned_{\datacenterindex, \attackindex,\vertexindexi,\vmindex,\aserver, \vertexindexj, \vmindex',\aserver',\trafficunit}=0$. Here $\trafficunit$ is an auxiliary subscript 
indicating that the one unit of traffic associated with $\vmassigned$ 
is the $\trafficunit$th one out of $\maxVol$ possible units of traffic.
The maximum required number of VMs of any type is denoted by 
$\maxVM$.

The ILP involves two key  decision variables: \emph{(1)} 
$\fraction_{\edgepopindex,\attackindex,\datacenterindex}$
is the fraction of traffic $\Traffic_{\edgepopindex,\attackindex}$ 
to send to datacenter $\datacenter_\datacenterindex$, and
\emph{(2)} $\vms_{\attackindex,\vertexindexi}^{\datacenterindex,\aserver}$
is the number of VMs of type 
$\vertex_{\attackindex,\vertexindexi}$ on server $\aserver$ of 
datacenter $\datacenterindex$, hence physical graphs
$\graph^{\physical}_\attackindex$.

\mypara{Objective function} The objective function (1) is composed 
of inter-datacenter and intra-datacenter costs, where constant $\alpha > 0$ 
reflects the relative importance of inter-datacenter cost to 
intra datacenter cost.

\mypara{Constraints} Equation (2) ensures all suspicious traffic will 
be sent to data
centers for processing. Equation (3) computes the amount of 
traffic of each attack type going to each datacenter, which is ensured
to be within datacenters bandwidth capacity using (4). Equation
(5) is intended to ensure sufficient numbers of VMs of the required
types in each datacenter. Servers compute capacities are enforced
using (6). Equation (7) sums up the cost associated with each 
datacenter, which is composed of two components: intra-rack cost,
given by (8), and inter-rack component, given by (9). Equation
(10) ensures the traffic processing capacity of each VM is not 
exceeded. Equations (11) and (12) tie the variables for number 
of VMs (i.e., 
$\vms_{\attackindex,\vertexindexi}^{\datacenterindex,\aserver}$)
and traffic (i.e., 
$\vmassigned_{\datacenterindex, \attackindex,\vertexindexi,\vmindex,\aserver, \vertexindexj, \vmindex',\aserver',\trafficunit}$)
to each other. Flow conservation of nodes is guaranteed by (13).
Inequality (14) ensures no ISP backbone link gets congested 
(i.e., by getting a traffic volume of more than a fixed fraction $\beta$ of its
maximum capacity), while 
$Path_{\edgepopindex \rightarrow \datacenterindex}$ is a path from 
a precomputed set of paths from $\edgepopindex$ to $\datacenterindex$.
The ILP decision variables are shown 
in (15).\comment{\footnote{There are two more constraints for sanity checking that we 
do not show here for brevity. The first one precludes the case of 
having both nodes of a graph edge from being realized 
on the same VM. The second sanity checking constraint ensures 
that there are no unnecessary VMs on the nodes of  a  
graph that have no incoming edges.}}

\comment{
As a simplifying assumption to scope the search space of the ILP solver, 
we explicitly formulate the problem such that the processing of a given 
end-to-end physical chain is within a single datacenter due to significantly 
higher values of cross datacenter latency compared to intra datacenter
latency in real world ISPs.}

\section{DSP and SSP Algorithms}
\label{subsec:dsp_ssp}

\vspace{-2mm}

As described in \Section\ref{sec:dist}, due to the impractically long time needed
to solve the ILP formulation, we design the DSP and SSP heuristics for resource 
management. The ISP global controller solves the DSP problem to assign suspicious 
incoming traffic to data centers. Then each local controller solves an SSP problem 
to assign servers to VMs. Figure~\ref{fig:dsp} and~\ref{fig:ssp} 
show the detailed pseudocode for the DSP and SSP heuristics, respectively.

\begin{figure}[h!]
\begin{minipage}{220pt}
 \fbox{
  \begin{minipage}{200pt}

{\footnotesize
\begin{codebox}

\li \Comment \emph{Inputs}: $\boldsymbol{\cost}$, $\boldsymbol{\Traffic}$, $\graph^{\logical}_{\attackindex}$, $\capacity^{link}_{\datacenterindex}$, and $\capacity^{compute}_{\datacenterindex}$

\li \Comment \emph{Outputs}: $\graph^{\physical}_{\attackindex,\datacenterindex}$
and
$\fraction_{\edgepopindex,\attackindex,\datacenterindex}$
values

\li 
  
\li Build max-heap  $\boldsymbol{\Traffic}^{maxHeap}$ of attack volumes $\boldsymbol{\Traffic}$ 
  
\li \While $!Empty(\boldsymbol{\Traffic}^{maxHeap})$
 
\li \Do $\atraffic \gets ExtractMax(\boldsymbol{\Traffic}^{maxHeap})$

\li $\datacenterindex \gets$ datacenter 
 with min. $\cost_{\atraffic.\edge,\atraffic.{\datacenterindex}}$ and cap.$>0$

\li \Comment enforcing datacenter link capacity 

\li $\atraffic_1 \gets min(\atraffic, \capacity^{link}_{\datacenterindex})$

\li  \Comment compute capacity of $\datacenterindex$ for traffic type $\attackindex$
 
\li $ \atraffic_2 \gets 
 \frac{\capacity^{Compute}_{\datacenterindex}}{\sum\limits_{\vertexindexi}{}
 \frac{\sum\limits_{\vertexindexj}{}{\weight_{\attackindex,\vertexindexj \rightarrow \vertexindexi}}}{\power_{\attackindex,\vertexindexi}}}$

\li  \Comment enforcing datacenter compute capacity 

\li $\atraffic_{assigned} \gets min(\atraffic_1, \atraffic_2)$ 
 
\li $\fraction_{\edgepopindex,\attackindex,\datacenterindex} \gets 
 \frac{\atraffic_{assigned}}{\traffic_{\atraffic.\edgepopindex,{\atraffic}.\attackindex}}$
 
\li \For each module type \vertexindexi

\li \Do \Comment update
 $\vms_{\attackindex,\vertexindexi}^\datacenterindex$ given new assignment

\li  $\vms_{\attackindex,\vertexindexi}^\datacenterindex = 
\vms_{\attackindex,\vertexindexi}^\datacenterindex +
 \atraffic_{assigned}^\datacenterindex  
 {
 \frac{\sum\limits_{\vertexindexj}{}{\weight_{\attackindex,\vertexindexj \rightarrow \vertexindexi}}}{\power_{\attackindex,\vertexindexi}}}$ 

\End

\li $\capacity^{link}_{\datacenterindex} \gets \capacity^{link}_{\datacenterindex} -  \atraffic_{assigned} $ 

\li $\capacity^{compute}_{\datacenterindex} \gets 
 \capacity^{compute}_{\datacenterindex} -  \atraffic_{assigned}  
 {\sum\limits_{\vertexindexi}{}
 \frac{\sum\limits_{\vertexindexj}{}{\weight_{\attackindex,\vertexindexj \rightarrow \vertexindexi}}}{\power_{\attackindex,\vertexindexi}}}$ 

\li  \Comment leftover traffic 

\li $\atraffic_{unassigned} = \atraffic - \atraffic_{assigned}$
 
\li \If ($\atraffic_{unassigned} > 0$) 
 
\li 		\Then $Insert(\boldsymbol{\Traffic}^{maxHeap}, \atraffic_{unassigned})$
\End

\End
 
\li \For each datacenter $\datacenterindex$ and attack type \attackindex
 
\li \Do Given $\vms_{\attackindex,\vertexindexi}^\datacenterindex$
and
$\graph^{\logical}_{\attackindex}$,
compute $\graph^{\physical}_{\attackindex,\datacenterindex}$

\End
 
\end{codebox}
}
\end{minipage}
}
\vspace{-3mm}
\caption{Heuristic for datacenter selection problem (DSP).}
\label{fig:dsp}
\vspace{-4mm}

\end{minipage}
\end{figure}

\hspace{0.2cm}
\begin{figure}[th!]
\vspace{-0.7cm}
\begin{minipage}{220pt}
  \fbox{
  \begin{minipage}{200pt}
  \footnotesize{
\begin{codebox}
\li \Comment \emph{Inputs}: $\graph^{\physical}_{\attackindex,\datacenterindex}$,
$\unitIntraRackCost$, $\unitInterRackCost$, 
\\and $\capacity^{compute}_{\datacenterindex,\aserver}$ values

\li \Comment \emph{Outputs}: $\vms_{\attackindex,\vertexindexi}^{\datacenterindex,\aserver}$  values

\li

\li \While entire $\graph^{\physical}_{\attackindex,\datacenterindex}$ 
is not assigned to $\datacenterindex$'s servers 

\li \Do $N \gets$ $\vertex_{\attackindex,\vertexindexi}^{\logical}$ whose all predecessors  
are assigned

\li \If ($N == NIL$)

\li \Then  $N \gets \vertex^{\logical}_\attackindex$ 
with max $\power_{\attackindex,\vertexindexi}$ 

\End

\li $localize$(nodes of $\graph^{\physical}_{\attackindex,\datacenterindex}$ corresponding to $N$) 

\End

\li

\li \Comment function $localize$ tries to assign all of its 
\\input physical nodes to the same server or rack

\li $localize$(inNodes)\{

\li assign all  inNodes to  emptiest server 

\li \If failed 

\li \Then assign all  inNodes to  emptiest rack 

\li \If failed

\li \Then split inNodes $\vertexes^{\physical}_\attackindex$ across racks

\End

\End

\li update $\vms_{\attackindex,\vertexindexi}^{\datacenterindex,\aserver}$  values

\li \}

\end{codebox}
}
  \end{minipage}
}
\vspace{-3mm}
  \caption{Heuristic for server selection problem (SSP) at datacenter $\datacenterindex$.}
	\label{fig:ssp}
\end{minipage}
\end{figure}

\end{document}